\documentclass[conference,compsoc]{IEEEtran}
\IEEEoverridecommandlockouts
\usepackage{cite}
\usepackage{amsmath,amssymb,amsfonts}
\usepackage{algorithmic}
\usepackage{graphicx}
\usepackage{textcomp}
\usepackage{xcolor}
\def\BibTeX{{\rm B\kern-.05em{\sc i\kern-.025em b}\kern-.08em
    T\kern-.1667em\lower.7ex\hbox{E}\kern-.125emX}}

\usepackage{dsfont}
\usepackage[nolist,nohyperlinks]{acronym}
\usepackage{soul}
\usepackage{tcolorbox}
\usepackage{tabularray}
\usepackage{hyperref}

\UseTblrLibrary{diagbox}

\acrodef{IMU}[IMU]{\emph{Inertial-Measurement-Unit}}
\acrodef{SMPL}[SMPL]{\emph{Skinned Multi-Person Linear Model}}
\acrodef{VAE}[VAE]{\emph{Variational Autoencoder}}
\acrodef{AE}[AE]{\emph{Autoencoder}}
\acrodef{KL}[KL]{\emph{Kullback-Leibler divergence}}
\acrodef{CVAE}[CVAE]{\emph{Conditional Variational Autoencoder}}
\acrodef{MR}[MR]{\emph{Mixed Reality}}
\acrodef{VR}[VR]{\emph{Virtual Reality}}
\acrodef{MFA}[MFA]{\emph{Montreal Forced Aligner}}
\acrodef{FACS}[FACS]{\emph{Facial Action Coding System}}
\acrodef{ASR}[ASR]{\emph{Automatic Speech Recognition}}
\acrodef{DMM}[DMM]{\emph{Deep Motion Masking}}
\acrodef{CNN}[CNN]{\emph{Convolutional Neural Network}}
\acrodef{MSE}[MSE]{\emph{Mean Squared Error}}
\acrodef{PCA}[PCA]{\emph{Principal component analysis}}
\acrodef{SVM}[SVM]{\emph{Support Vector Machine}}
\acrodef{RBF}[RBF]{\emph{Radial basis function}}
\acrodef{ADE}[ADE]{\emph{Average displacement error}}
\acrodef{FDE}[FDE]{\emph{Final displacement error}}
\acrodef{LSTM}[LSTM]{\emph{Long Short-Term Memory}}
\acrodef{ReLu}[ReLu]{\emph{Rectified Linear Unit}}
\acrodef{EYKT}[EYKT]{\emph{Eye You Know Too}}

\newif\ifanon
\anonfalse

\newif\ifcomments
\commentstrue
\ifcomments
\newcommand{\ts}[1]{\textcolor{blue}{\textbf{TS:} #1}}
\newcommand{\adc}[1]{\textcolor{purple}{\textbf{AC:} #1}}
\newcommand{\sh}[1]{\textcolor{olive}{\textbf{SH:} #1}}

\newcommand{\old}[1]{\textcolor{red}{#1}}
\else
\newcommand{\ts}[1]{}
\newcommand{\adc}[1]{}
\newcommand{\sh}[1]{}

\newcommand{\old}[1]{}
\fi

\makeatletter
\newcommand{\linebreakand}{
  \end{@IEEEauthorhalign}
  \hfill\mbox{}\par
 \mbox{}\hfill\begin{@IEEEauthorhalign}
}
\makeatother

\begin{document}
\title{FacialMotionID: Identifying Users of Mixed Reality Headsets using Abstract Facial Motion Representations}
\ifanon
\author{}
\else
\title{FacialMotionID: Identifying Users of Mixed Reality Headsets using Abstract Facial Motion Representations}

\author{\IEEEauthorblockN{1\textsuperscript{st} Adriano Castro}
\IEEEauthorblockA{\textit{KASTEL Security Research Labs} \\
\textit{Karlsruhe Institute of Technology}\\
Karlsruhe, Germany \\
adriano.castro@student.kit.edu}
\and
\IEEEauthorblockN{2\textsuperscript{nd} Simon Hanisch}
\IEEEauthorblockA{\textit{Centre for Tactile Internet (CeTI)} \\
\textit{Technical University Dresden}\\
Dresden, Germany \\
simon.hanisch@tu-dresden.de}
\and
\IEEEauthorblockN{3\textsuperscript{rd} Matin Fallahi}
\IEEEauthorblockA{\textit{KASTEL Security Research Labs} \\
\textit{Karlsruhe Institute of Technology}\\
Karlsruhe, Germany \\
matin.fallahi@kit.edu}
\linebreakand 
\IEEEauthorblockN{4\textsuperscript{th} Thorsten Strufe}
\IEEEauthorblockA{\textit{KASTEL Security Research Labs} \\
\textit{Karlsruhe Institute of Technology}\\
Karlsruhe, Germany \\
thorsten.strufe@kit.edu}
}
\fi

\maketitle

\begin{abstract}

Facial motion capture in mixed reality headsets enables real-time avatar animation, allowing users to convey non-verbal cues during virtual interactions. However, as facial motion data constitutes a behavioral biometric, its use raises novel privacy concerns. With mixed reality systems becoming more immersive and widespread, understanding whether face motion data can lead to user identification or inference of sensitive attributes is increasingly important. 

To address this, we conducted a study with 116 participants using three types of headsets across three sessions, collecting facial, eye, and head motion data during verbal and non-verbal tasks. The data used is not raw video, but rather, abstract representations that are used to animate digital avatars. Our analysis shows that individuals can be re-identified from this data with up to 98\% balanced accuracy, are even identifiable across device types, and that emotional states can be inferred with up to 86\% accuracy. These results underscore the potential privacy risks inherent in face motion tracking in mixed reality environments.

\end{abstract}


\section{Introduction} \label{sec:intro}

\begin{figure}[!ht]
    \includegraphics[width=\linewidth]{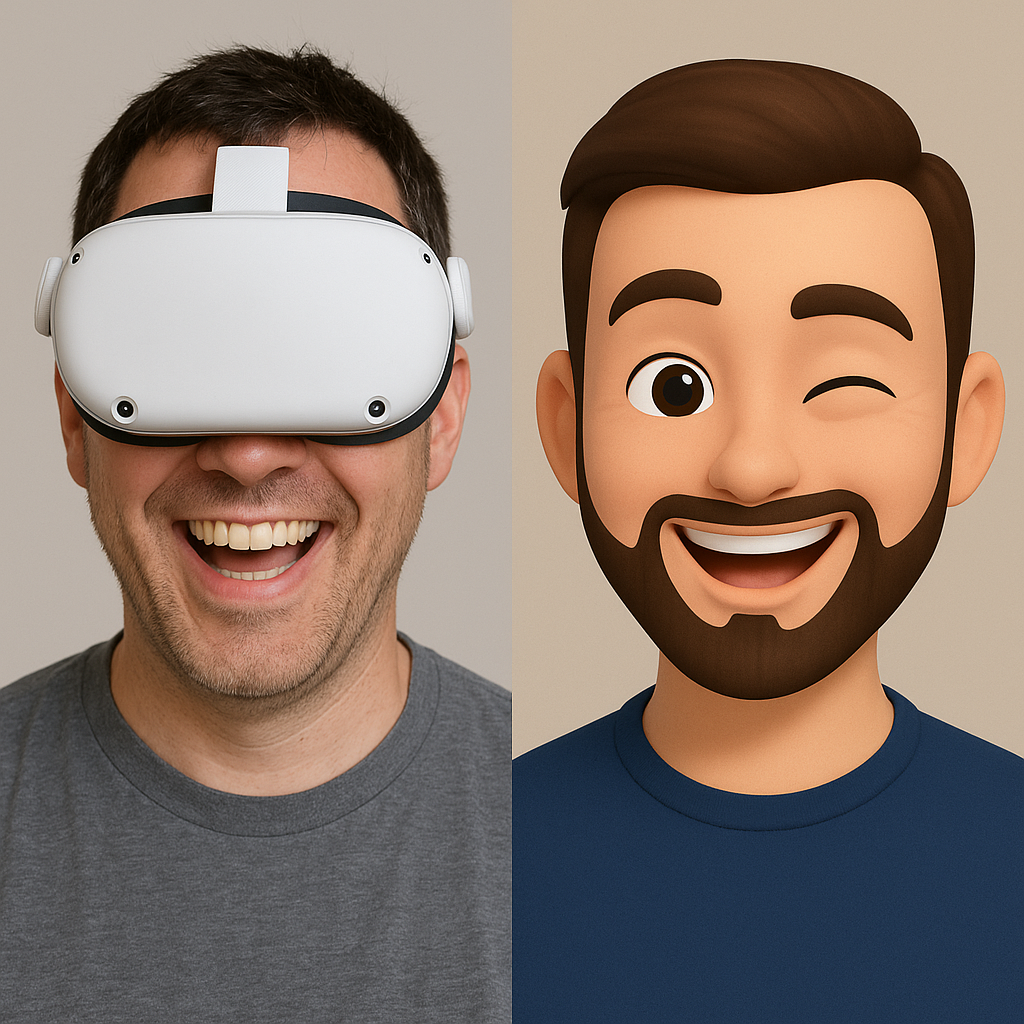}
    \caption{A user wearing a mixed reality headset with facial motion tracking. Their avatar mimics their facial expressions.}\label{fig:facial_motion_tracking}
\end{figure}

\ac{MR} promises to fuse the real and digital worlds. 
This implies the universal tracking of \ac{MR} users to create precise digital twins of them. Their appearance, voice, and motions are captured and streamed onto digital avatars. The newest generation of \ac{MR} headsets (e.g., Apple Vision Pro\footnote{https://www.apple.com/apple-vision-pro/} and Meta Quest Pro\footnote{https://www.meta.com/de/en/quest/quest-pro/}) already integrate face and eye tracking to animate the faces of these digital avatars (see Figure~\ref{fig:facial_motion_tracking} for an example). Integrating facial and eye motions improves social interactions in \ac{MR}, as subtle non-verbal cues can now be transmitted to a dialogue partner. Currently, we are still in the early adopter stage of this technology as only a handful of applications such as VRChat\footnote{https://hello.vrchat.com/} or virtual YouTubing (using a virtual character to create videos) make use of facial motions. Nonetheless, with the advancement of \ac{MR}, facial motion tracking is expected to become a standard feature of future \ac{MR} devices.

However, sharing facial motion data in \ac{MR} poses a potential privacy risk because facial motions are a behavioral biometric trait. 
It may yield both identity and attribute disclosure risks:
An attacker could use the facial motion data from the avatar shared in \ac{MR} to perform privacy inferences like identification or employ attribute inferences, like emotion recognition.

Imagine a user visiting a digital store in the Metaverse wearing an \ac{MR} headset. The user has a generic avatar that does not reveal their identity, and the avatar's facial motion tracking is turned on by default. Without the user's knowledge, the store owner can collect their facial motion data by observing the avatar's facial animations. The store owner can use this data to identify the user, determine if they have visited the store before, and recognize their facial expressions to see which items they like. Thus, the user shares much more private information than they realize.

Although many behavioral biometric traits, such as gait~\cite{hanisch2022understanding}, voice~\cite{cheng2022personal}, and eye gaze~\cite{kroger2020does}, are already known to be privacy sensitive, this remains an open question for facial motions.
Therefore, we seek to understand whether individuals can be identified from facial motion data and whether emotional states can be inferred. To this end, we designed and conducted a study in which we recorded 116 participants using three types of \ac{MR} headsets. Each participant attended a maximum of three sessions, with each session being approximately a week apart. During each session, participants were recorded with two types of headsets while performing a set of verbal and non-verbal tasks, with multiple repetitions of each task. 

We then performed privacy experiments using the collected dataset by investigating the general identification using three different biometric recognition models. Besides general identifiability, we investigate if participants can be re-identified across sessions and different \ac{MR} headset types. Further, we look at emotion recognition and examine which tasks are best suited for identification.

The main contributions of this paper are as follows:

\begin{itemize}
    \item We recorded a novel facial motion dataset, which for the first time allows the investigation of associated privacy risks.
    \item We demonstrate that the identification of individuals is possible from facial motion data alone.
    \item We show that re-identifying people across sessions and different \ac{MR} headsets is possible.
    \item We confirm that emotion recognition from abstract facial motion data can be performed with high accuracy.
\end{itemize}

The paper is organized as follows. In Section~\ref{sec:related_work}, we explore the related work of facial motion identification, and then describe the background in Section~\ref{sec:background}. We then first describe the general study design in Section~\ref{sec:study_design} before we describe our concrete study implementation in Section~\ref{sec:study_implementation}. Afterwards, we evaluate the study by performing identification experiments in Section~\ref{sec:evaluation}, and subsequently discuss them in Section~\ref{sec:discussion}. We end the paper with a short conclusion in Section~\ref{sec:conclusion}.

\section{Related Work}\label{sec:related_work}

In the following section, we will present the related research on identifying individuals in \ac{MR} through analysis of facial motion. 
The primary focus of our research lies in the development and analysis of methodologies for the identification of facial motion from video data, the detection of facial expressions, the identification through eye gaze, and the identification of individuals from \ac{MR} motion data.

\subsection{Facial Motion Identification}\label{sec:facial_identification}

Some preliminary studies have been conducted on the identification of individuals based on facial motion, with the majority of these studies focusing on video data.

Benedikt et al.~\cite{Benedikt_2010} employed 3D videos of faces to assess the distinctiveness of facial motion for biometric authentication. 
The trajectory of these facial motions is then represented within the Eigenvector space of diverse facial expressions. 
Their findings indicate that non-verbal tasks may not be as effective in terms of identification from facial motions as verbal tasks. 
Zhang et al.~\cite{10.1016/j.sigpro.2019.02.025} performed a similar study, in which they collected 3D videos of participants speaking a passcode 10 times. The system demonstrates an impressive capacity to identify the participant from the dynamic features of the video, achieving a 96\% accuracy rate with 77 participants. Haamer et al.~\cite{Haamer_2018} collected a video dataset of 61 participants performing various emotion tasks. They then show that participants can be identified using the videos recorded.

Moreira et al.~\cite{Moreira_2022} utilized a neuromorphic sensor, an advanced device capable of capturing precise alterations in individual pixels, to record the facial expressions of 40 participants while reciting nursery rhymes. They can show that identification is possible with accuracies as high as 96\%.

The existing literature suggests that the identification of individuals through facial motion is feasible for both facial expression and speaking tasks. However, given that the majority of studies employ video data, it remains uncertain whether identification can be achieved exclusively through the analysis of facial movements alone, since face recognition is possible on static face images. Additionally, the question remains open whether individuals can be identified across multiple sessions via facial motion data.

\subsection{Facial Expression Recognition}\label{sec:facial_exp_recog}

One field of study that has focused on facial motion analysis is facial expression recognition. The objective of facial expression recognition is to categorize the emotions displayed by the individual captured on video~\cite{Kopalidis_2024}. Zhao et al.~\cite{Zhao_2021} propose a lightweight model to extract the displayed emotion from face images. Wen et al.~\cite{Wen_2023} use an attention network to perform emotion recognition and achieve state-of-the-art performance. Furthermore, Chen et al.~\cite{Chen_2019} have employed the differences between a neutral face and an expressive face to enhance the learning of different face expressions. To improve generalization in their face recognition model, Zhang et al.~\cite{Zhang_2021} propose learning an identity-independent representation of facial expressions using deviation learning. This involves subtracting a person's identity, established by a face recognition model, from their facial expression embedding.

Lee et al.~\cite{Lee_2024} investigate facial expression recognition using a face mask that measures facial deformation, rather than via videos.

Facial expression recognition has also already been investigated in the context of \ac{MR} by Chen et al.~\cite{Chen_2022} in a study in which extra cameras have been integrated into an existing \ac{MR} headset. Additionally they used an external camera to capture the part of the face which is not hidden behind the MR headset. They then show that they can achieve a facial expression recognition accuracy of 95\%.

Facial expression recognition shows that facial motion data is useful for more than just direct social interactions between people. However, this information should also be considered private, and individuals should have the choice of when and how they share their emotions.

\subsection{Eye Gaze}\label{sec:eye_gaze_identification}

Eye gaze was recognized as a privacy-sensitive topic some time ago and has also drawn attention as a possible behavioral biometric trait for authentication. Lohr et al.~\cite{lohr_metric_2020} showed that they could identify 269 subjects with a mean EER of 4.72\% using the SBA-ST dataset~\cite{Friedman_2017}, which was captured with a dedicated eye tracker. They further improved their method in EyeKnowYouToo~\cite{9865991}, which is the current state-of-the-art model for user authentication based on eye gaze. They achieved an EER of 3.66\% at a sampling rate of 1000 Hz and an EER of 8.77\% at a sampling rate of 125 Hz. In a later study, Raju et al.~\cite{raju2024evaluatingeyemovementbiometrics} investigated the performance of eye gaze authentication on the GazeBasedVR~\cite{Lohr_2023} dataset and showed that short-time authentication works well but that the EER increases to 10\% for longer sessions.

Shao et al.~\cite{shao_cross-content_2024} aim to create an eye-gaze identification system in \ac{MR} that is independent of the content shown to users. They use two encoders: one for content and one for eye gaze. They achieved an F-score of 92\%. Asish et al.~\cite{Asish_2022} use eye gaze features of 34 people performing four different tasks for identification in \ac{VR}.

As the privacy-sensitive nature of eye gaze data has been recognized, the first studies~\cite{Liu_2019, Ren_2024, 10464194} seeking to anonymize it have emerged. Common methods of anonymization include adding noise or smoothing the eye gaze trajectories.

Eye gaze is useful not only for authentication, but also for foveal rendering. Foveal rendering is a selective rendering process that increases the level of detail in the section of the image at which the user is looking. Several studies~\cite{Arabadzhiyska_2017, Hu_2020, Ding_2025} attempt to predict eye gaze to enable foveal rendering.

The research on eye gaze data showcases the dual nature of behavioral biometric data, as both privacy inferences, as well as desired applications like authentication and foveal rendering are possible with it.

\subsection{Mixed Reality Identification}

In recent years, the subject of identifying people using motion data recorded by \ac{MR} headsets has gained traction, and multiple studies have been published on the topic. Among the first of these studies, Miller et al.~\cite{Miller_2020} recorded 511 participants watching 360-degree videos in \ac{VR}. The researchers demonstrated a high identification rate of 95\% using the head and controller motions. Liebers et al.~\cite{Liebers_2021} demonstrated that identifying individuals is possible by combining the head orientation and eye gaze of 12 people captured with a \ac{MR} headset. Moore et al.~\cite{Moore_2021} investigated which \ac{VR} tasks are most effective for identification, once again using headset and controller motions. They found that identification success depends on the \ac{VR} content used. Nair et al.~\cite{nair2023unique} used a large-scale dataset of people playing Beat Saber\footnote{https://www.beatsaber.com/} and demonstrated their ability to identify players in a pool of over 50,000 people with 94\% accuracy using 100 seconds of headset and controller motion data.

\section{Background}\label{sec:background}

Here, we briefly describe the background for \ac{MR} motion tracking and biometric recognition required for this work.

\begin{figure}[!bt]
    \includegraphics[width=\linewidth]{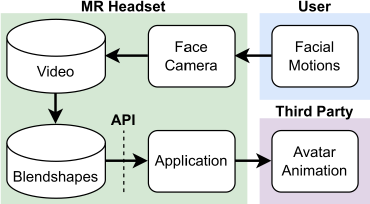}
    \caption{The data sharing pipeline of facial motion data captured by \ac{MR} headsets.}\label{fig:facial_motion_data_sharing}
\end{figure}

\begin{figure}[!bt]
    \includegraphics[width=\linewidth]{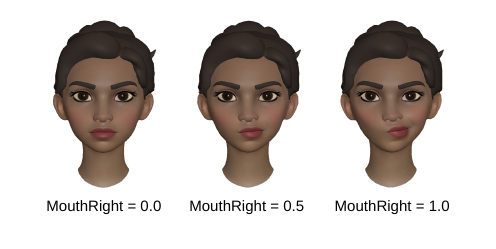}
    \caption{A blendshape named ``MouthRight'' being activated on an \ac{MR} avatar from 0 to 1 through interpolation.}\label{fig:blendshapes}
\end{figure}

\subsection{Mixed Reality Tracking}

\textbf{Facial Motion Tracking:} The \ac{MR} headsets used in our study rely on camera-based face tracking. 
Inward-facing infrared cameras capture the eyes and mouth of the person wearing the headset. 
This video data is then transformed into a symbolic representation which is shared via applications on the \ac{MR} headset. See Figure~\ref{fig:facial_motion_data_sharing} for the full data sharing pipeline of facial motion data.
For facial motions, the data is represented as blendshapes. Blendshapes are a type of interpolated animation, also known as morph target animation. In this type of animation, the neutral state and deformed version of an object are stored for each blendshape. Then, for each frame of the animation, the object's vertices are interpolated between the neutral and deformed versions. An example of a blendshape for an \ac{MR} avatar is the right part of the mouth (see Figure~\ref{fig:blendshapes}). In the neutral state, the mouth is symmetrical; in the deformed state, it is pulled to the right side of the face. All intermediate states can be created via interpolation. The blendshapes defined by the \ac{MR} headsets are usually based on the \ac{FACS}~\cite{ekman_facs_2024, meta_facs}. The former is a system that defines and describes all distinguishable facial movements, so-called action units. These action units are derived from anatomy, and with them complete expressions can be recognized objectively. Two examples of such action units are ``Cheek Raiser'' and ``Lip Corner Puller'' that together can be interpreted as the expression of happiness \cite{cmu_facs_2001}.

\textbf{Eye Tracking:} The user's eye gaze is captured via infrared cameras positioned inside the \ac{MR} headset. The video is then converted into gaze direction and eye position data.

\textbf{Motion Tracking:} For the motion tracking in \ac{MR} headsets there exist two main approaches. Inside-out tracking describes the approach in which multiple cameras on the outside of the headset are used to establish its position. The second approach is light house tracking in which one or multiple static light houses emit sequences of infrared light which are registered by infrared sensors on the surface of the headsets and the controllers. The headset and controllers can then compute their distance and orientation in relation to the light houses. When comparing the two approaches, inside-out tracking is less precise but easier to use than lighthouse tracking.

\subsection{Biometric Recognition}

\textbf{Biometric traits} (sometimes \textbf{biometric characteristics}~\cite{ISO2382}) are properties of a human that either capture what a human is (e.g. face, iris) or how a human behaves (e.g. voice, gait, heartbeat).
The former are known as biological, the latter as behavioral biometric traits. 

\textbf{Biometric recognition} is the process of inferring the identity or specific attributes of an individual from its biometric data. For inferring the identity we consider two cases. 
\textbf{Authentication} entails the verification of a claimed identity given the input of one fresh observation and a template representing the class of that claimed identity. 
The main threat to biometric authentication is impersonation: An adversary succeeding a verification attempt for another individual's identity.
\textbf{Identification} of a given observation produces the most likely candidate (or: list of top-k candidates) from all learned classes that represent individuals, together with their respective classification confidence.
Complementary to such identity disclosure, \textbf{Attribute inference} is a privacy threat in which a specific private attribute (e.g., age, sex, medical condition) of an individual is inferred from the biometric data.

\section{Study Design}\label{sec:study_design}

In this section, we describe the design of our study to investigate identity and attribute disclosures from abstract facial motion data. We first explain the general rationale before providing a more detailed explanation of the tasks used and the selected recording schedule.

\subsection{Design Rationale}

The main goal of our study is to investigate whether identifying individuals from their facial motion data is possible.
To allow biometric recognition systems to train on the data and recognize identifying patterns, we require a large number of samples. 
Therefore, we require numerous repetitions and task executions involving a diverse group of participants.
Additionally, we aim to determine whether facial motion data is a stable biometric factor over time; therefore, we will record multiple sessions with each participant. 
Lastly, we want to investigate whether facial motion data generalizes well when different devices are used to capture facial expressions. 
Therefore, we record our participants using multiple device types that integrate facial motion tracking.

We see the main application of facial motion data for animating digital avatars as speaking to other people and displaying emotions. Consequently, we focus on tasks involving two types of categories, namely speech and emotional expression for data collection. 
As mentioned in Section~\ref{sec:facial_exp_recog}, emotion recognition has been shown to work previously. 
Hence, we integrate it into the study to test collected data and to compare results. 
Since facial motion data will likely be used in combination with eye gaze and head motion data---and as these are readily available in the common MR headsets---we also collect these.

\subsection{Recording Procedure}

We chose to record our participants over the course of three separate sessions, with each session being approximately a week apart from each other. 
In the first session, participants first answer a short questionnaire about demographics before the actual recording starts. 
During each session, we record each participant performing the same set of tasks with two different \ac{MR} headset types. 
We chose to keep one headset type the same throughout all sessions, whereas the respective other headset was alternated between the remaining two in each session. This allowed us to record all participants using three different headsets. Due to the change in the second headset, we split our participants into two groups, A and B, to keep track of which second headset had to be used in each session.

\subsection{Tasks}

We designed a task-based study in which participants performed predefined tasks sequentially. An overview can be seen in Table~\ref{tab:tasks}. At the beginning of the study, one tutorial task was performed for each task type.
To cover the described applications, we selected verbal tasks, in which participants read a given text aloud, and non-verbal tasks, in which participants mimic a facial expression. Studies such as ~\cite{Benedikt_2010, Moreira_2022} have demonstrated that verbal tasks contain the most identity cues in facial motion, unlike non-verbal tasks. Therefore, the predominant task category we selected is verbal tasks.

First, the participant is shown the current task.
Then, the participant starts the actual recording phase for the task by pressing a button.
During the recording phase, the participant performs the task.
The recording phase is ended by pressing the same button again.
All tasks and their repetitions are presented to the participant in a random order.
There are four repetitions for each task in the first session and five repetitions for each task in the second and third sessions.
The reduction of repetitions in the first session allows time for the questionnaire.

\begin{figure}[!tbp]
  \centering
  \begin{minipage}[b]{0.235\textwidth}
    \includegraphics[width=\textwidth]{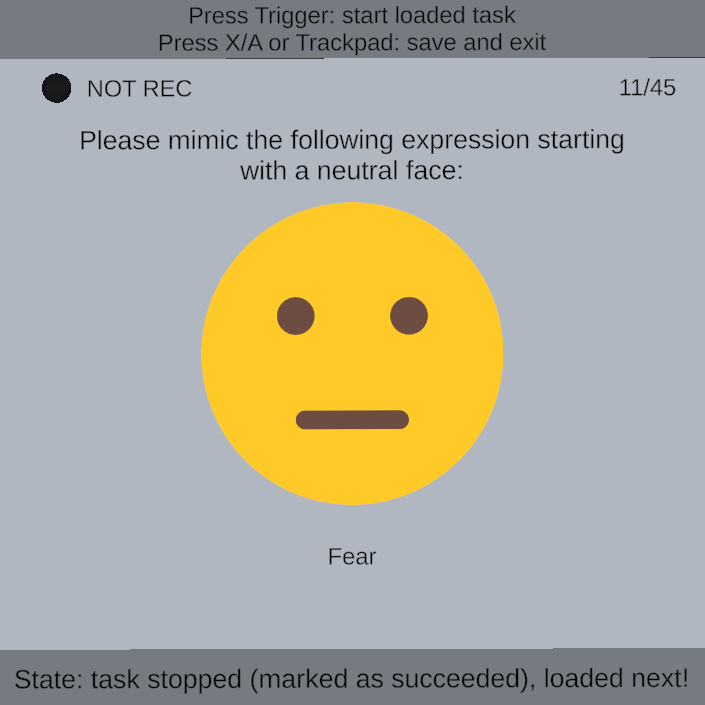}
    \caption{The participant performs the expression starting with a neutral face after pressing the button.}\label{fig:nonverbal_task_1}
  \end{minipage}
  \hfill
  \begin{minipage}[b]{0.235\textwidth}
    \includegraphics[width=\textwidth]{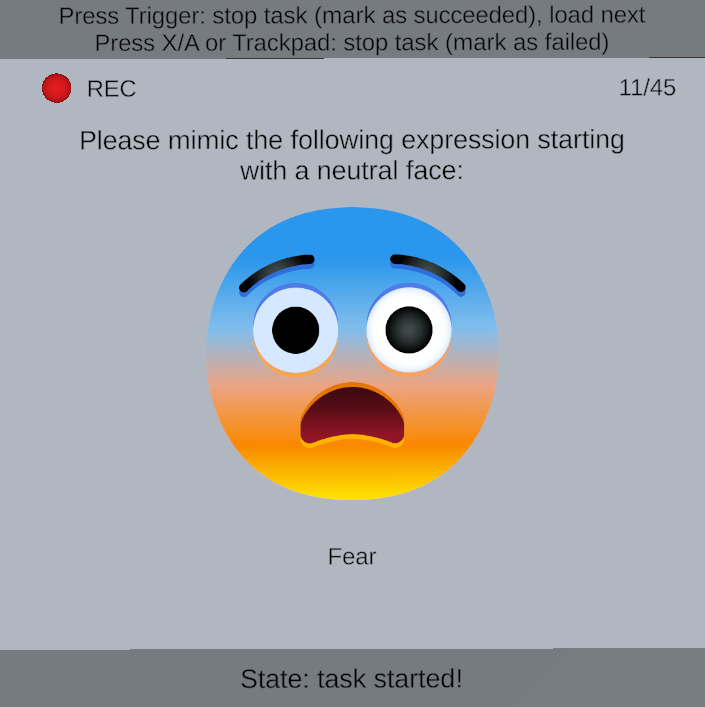}
    \caption{The participant performs the expression starting with a neutral face after pressing the button.}\label{fig:nonverbal_task_2}
  \end{minipage}
\end{figure}

\textbf{Non-verbal Tasks:}
We presented the non-verbal tasks using emoticons that displaying three different facial expressions: happiness, anger, and fear. See Figure~\ref{fig:nonverbal_task_1}+\ref{fig:nonverbal_task_2} as an example for a non-verbal task.
This abstract representation should encourage participants to perform the facial expressions as they normally would rather than closely mimicking the avatars shown to them. Therefore, we did not use high-fidelity digital avatars. 
We instructed participants to mimic the non-verbal tasks shown to them by starting with a neural facial expression and to then transition into the shown facial expression.
An animation of the emoticon changing from neutral to the target expression illustrates this process.

\begin{figure}[!ht]
    \centering
    \includegraphics[width=0.7\linewidth]{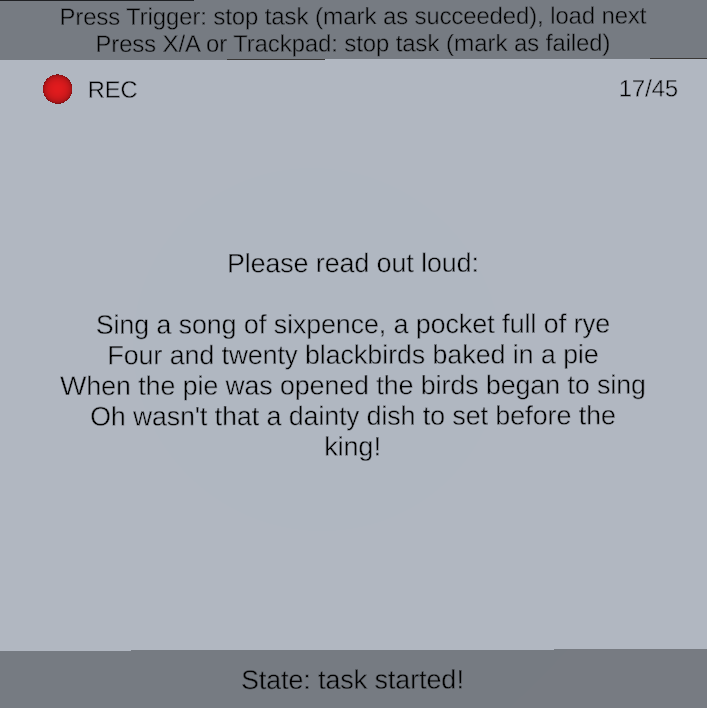}
    \caption{An example of a verbal task in which the participant is uttering the nursery rhyme ``Sing a Song of Sixpence''.}
    \label{fig:verbal_task}
\end{figure}

\textbf{Verbal Tasks:} During the verbal tasks (see Figure~\ref{fig:verbal_task}), participants are asked to utter words and sentences.
Lu et al.~\cite{lu_lippass_2018} have shown that words and groups of sentences that contain a large number of phonemes are best suited for identification.
A phoneme is the smallest unit of sound which makes a lexical difference in a language.
Additionally, Moreira et al.~\cite{Moreira_2022} have already shown that reciting nursery rhymes are suitable for facial motion identification. Therefore, we selected nursery rhymes for the verbal tasks because they contain various repetitive phonemes. To select the nursery rhymes, we used a list\footnote{https://www.bbc.co.uk/teach/school-radio/articles/z4ddgwx} of common English nursery rhymes. To keep the verbal task short, we prepared the list by splitting all rhymes, such that each part is at most four lines long.
Next, we counted the phonemes of each nursery rhymes and selected the top three with the highest count. Out of these selected nursery rhymes, we selected one word of each that contained the highest amount of phonemes, constituting the word tasks.

\begin{table}[h]
\centering
\caption{Overview of the different tasks which the participants performed in the study. v: verbal, nv: non-verbal}\label{tab:tasks}
\begin{tabular}{|l|l|l|l|} \hline
    \textbf{ID}  & \textbf{Type}       & \textbf{Task} & \textbf{Repetitions} \\ \hline
    0   & v         & sixpence (word)                   & 4 / 5         \\ \hline
    1   & v         & dinosaurs (word)                  & 4 / 5          \\ \hline
    2   & v         & muffin (word)                     & 4 / 5          \\ \hline
    3   & v         & Sing a Song of Sixpence (rnhyme)  & 4 / 5          \\ \hline
    4   & v         & Dinosaurs (nrhyme)                & 4 / 5          \\ \hline
    5   & v         & The Muffin Man (nrhyme)           & 4 / 5          \\ \hline
    6   & nv        & happiness                         & 4 / 5          \\ \hline
    7   & nv        & anger                             & 4 / 5          \\ \hline
    8   & nv        & fear                              & 4 / 5          \\ \hline
\end{tabular}

\end{table}

\section{Study Implementation}\label{sec:study_implementation}

The study was conducted between January 22 and February 14, 2025.
It took place in a dedicated laboratory that contains multiple small booths specifically designed for user studies, and an office for their supervision.
We divided each study day into 12 slots, with each day ranging from 8:30 am to 6:15 pm.
Since we aimed for a study duration of approximately 30 minutes, an equal time allocation was assigned to each slot.
To compensate for unexpected duration times, we added a 15-minute break between each slot.
At each slot, two individuals participated simultaneously --- one from group A and another from group B.
As each booth contained a door, each participant could perform the study without any disturbances.

\subsection{Ethics}

The data collection was approved by the ethics
commission of \ifanon our research institution \else the Karlsruhe Institute of Technology (research project ”Privacy of Facial Motions”)\fi and was
conducted in accordance with the Declaration of Helsinki. Participants were paid based on their time of participation at an hourly rate of 14€. Additionally, participants received a flat bonus of 2€ or 3€ for participating in the second and third sessions, respectively. We obtained informed consent from all participants for the data collection and processing.

\subsection{Apparatus}

During the study, we used four \ac{MR} devices, namely two Meta Quest Pros, one Pico 4 Enterprise\footnote{https://www.picoxr.com/global/products/pico4e}, and one HTC Vive Pro Eye\footnote{https://www.vive.com/sea/product/vive-pro-eye/overview/} with the Facial Tracker add-on\footnote{https://developer.vive.com/us/hardware/facial-tracker/}. All of these devices support eye and facial tracking in addition to standard head and controller tracking. 
Moreover, the devices and their tracking are supported by Unity, the Game Engine that we used to implement the application for our study. 
While the first device type is designed for both augmented and virtual reality, the other two are purely \ac{VR} devices. 
Since we only require \ac{VR}, the three types of devices were deemed suitable for our experiments.

The study was implemented as a Unity application since all selected \ac{MR} devices supported it.
Unity Engine v2021.3.32f1 was utilized for development, as it was the most recent long-term support version supported by all headsets and their tracking APIs.
We created a scene for each device, as they required individually configured XR cameras and device specific code to activate their motion tracking.

To be able to access the motion data of the devices and store them, we utilized several Unity packages that allowed the interaction with the \textbf{APIs} of the devices. 
For the Meta Quest Pro we used the Meta Movement SDK v71.0.1 including the Meta XR Core v71.0.0 and the Meta XR Interaction SDKs v71.0.0\footnote{https://developers.meta.com/horizon/documentation/unity/move-overview/}. 
For the Pico 4 Enterprise we used the PICO Unity Integration SDK v2.5.0\footnote{https://developer.picoxr.com/document/unity/?v=2.5.0}. 
And for the HTC Vive Pro Eye we used the VIVE OpenXR Plugin v2.0.0\footnote{https://github.com/ViveSoftware/VIVE-OpenXR-Unity} with addition of the VIVE SRanipalRuntime v1.3.1.1 and the OpenXR Plugin v1.9.1 for the facial tracker.
Both our Meta Quest Pros used during the study had identical software and runtime as well as OS versions, namely v71.0.0 and SQ3A.220605.009.A1 respectively.
The Pico 4 Enterprise ran on version v5.9.9, and the Vive's eye and lip camera versions were v2.41.0-942.e3e4 and v50100 in corresponding order.

\subsection{Recruitment}

We recruited 116 participants (45 female, 71 male; age mean 23.6 years, std 4) with the help of \ifanon a panel \else the KD2Lab panel \fi of \ifanon our institution. \else the Karlsruhe Institute of Technology.\fi
The distance between two subsequence sessions was between 4-16 days (participants per session 1: 116, session 2: 83, session 3: 49). Of the participants, 67 were native German speakers, while the rest reported a different mother tongue. 67 describe themselves as ambiverts, 26 as extroverts, and the remaining 23 as introverts.

The participants were assigned to their respective group at random. While group A used the HTC Vive Pro Eye in addition to their assigned Meta Quest Pro in the first session, group B started with the Pico 4 Enterprise.
In the second session, group A then received the Pico 4 Enterprise instead of the HTC Vive Pro Eye, and group B vice versa.
In the third session, group A and B each returned to their first headsets.
Thus, each participant who participated in all sessions used each device at least once and the Meta Quest Pro three times.

\subsection{Session Procedure}

\begin{figure}[!ht]
    \includegraphics[width=\linewidth]{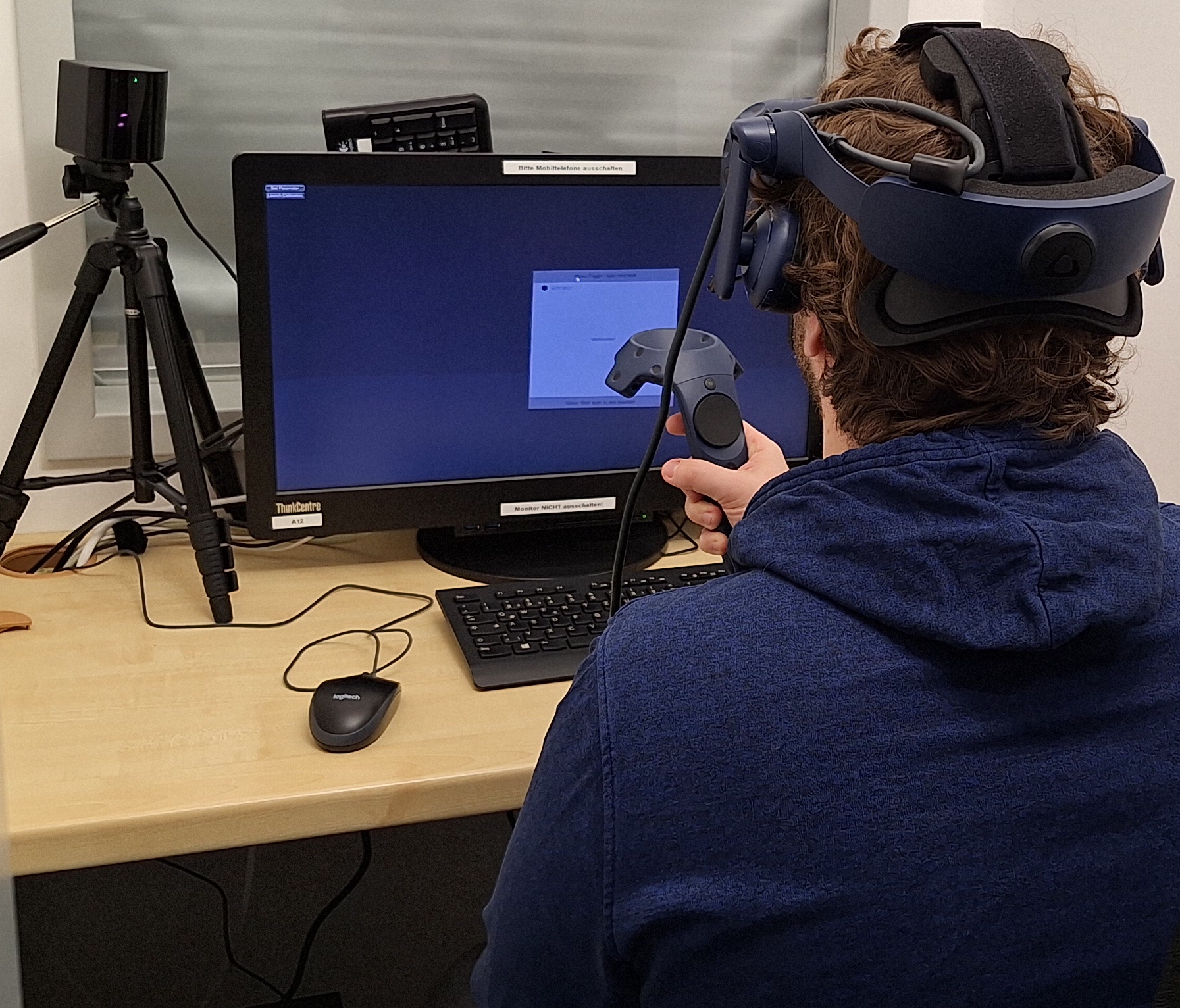}
    \caption{A participant performing the tasks with the HTC Vive Pro Eye.}\label{fig:example_participant}
    \label{fig:action_vs_recognition}
\end{figure}

For the first session, our participants required more thorough guidance and support.
We began by introducing the study and explaining the procedure, emphasizing the data collection process and its purpose. Then, we started a timer to keep track of their study duration, which was relevant for their payment at the end. Then, we assigned each participant a random pseudonym to be used for the remainder of the study.

Next, we escorted each participant to their assigned room. Each participant was given an information sheet with details about the study, a data protection agreement, and a survey. The survey collected information about the participants' age, sex, origin, self-assessed personality traits, English proficiency, and mother tongue. After completing the survey, the participants watched a short introductory video showing them how to use the MR headsets and their respective calibration procedures.

After watching the tutorial videos, the participants were brought to the booth with their first headset.
We helped them become accustomed to the headset and to perform the eye calibration.
Thereafter, the participants started the Unity application and, thus, performed the tasks shown through their \ac{MR} headset.
When they completed the tasks with the first headset, they were brought to the second one, where we repeated the procedure.
At the end, the participants filled in a short online survey to receive their payment with their own payout token assigned through the experiment organization.
To reduce any possible bias in the data due to headset order, the order of the headsets was inverted for each group of participants. See Figure~\ref{fig:example_participant} for an example how the participants performed the study.

In subsequent sessions, participants did not have to fill out the survey or data protection sheet again.
Although we asked the participants if they wanted to watch the eye calibration tutorial videos again, they usually skipped them since they remembered how to perform the tasks.
Additionally, the subjects usually skipped reading the study information sheet from the first session.
They were usually brought directly to the headsets and performed the study as described above.

\subsection{Troubleshooting}

During the study, there were some difficulties.
For the first recording day (22.01.2025) we encountered a problem for the facial motion recording of the HTC Vive, and as a consequence the HTC Vive recordings for the first day contain less blendshapes then the following recordings.
Another problem we encountered with the HTC Vive was that for some of the audio recordings the recording frequency was higher than configured, though this was unproblematic since our data processing approach presented in Section~\ref{sec:data_proc} is robust against it.
The eye calibration of the Meta Quest Pro devices turned out to be challenging, as it would regularly finish unsuccessfully.
This seemed to be more frequent with participants wearing glasses, yet it also happened with non-glasses wearers.
In such problematic cases, we helped the participants adjust the lenses and the position of the headset on their heads --- it did help a relative number of cases, but not all of them. Due to these problems, the quality of the eye tracking for the Meta Quest Pro suffered.
Another challenge was that both the eye and face tracking of the Meta Quest Pro devices tended to suddenly malfunction in between participants.
This happened once per Meta Quest Pro device, and was unfortunately only discovered at the end of the day. Due to this issue, we lost 19 recordings.

\subsection{Data Processing}\label{sec:data_proc}

\providecommand{\ViveFaceCheekPuff}{Cheek\_Puff\_*}
\providecommand{\PicoFaceCheekPuff}{CheekPuff}
\providecommand{\MetaFaceCheekPuff}{CheekPuff*}
\providecommand{\CommentFaceCheekPuff}{left/right}

\providecommand{\ViveFaceEyeClosed}{Eye\_*\_Blink}
\providecommand{\PicoFaceEyeClosed}{EyeBlink\_*}
\providecommand{\MetaFaceEyeClosed}{EyesClosed*}
\providecommand{\CommentFaceEyeClosed}{left/right}

\providecommand{\ViveFaceEyeLook}{Eye\_*\_*}
\providecommand{\PicoFaceEyeLook}{EyeLook*}
\providecommand{\MetaFaceEyeLook}{EyesLook*}
\providecommand{\CommentFaceEyeLook}{left/right, down/up, in/out}

\providecommand{\ViveFaceJaw}{Jaw\_*}
\providecommand{\PicoFaceJaw}{Jaw*}
\providecommand{\MetaFaceJaw}{Jaw*}
\providecommand{\CommentFaceJaw}{forward/thrust, left/right, open/drop}

\providecommand{\ViveFaceLidTightener}{Eye\_*\_Squeeze}
\providecommand{\PicoFaceLidTightener}{EyeSquint\_*}
\providecommand{\MetaFaceLidTightener}{LidTightener*}
\providecommand{\CommentFaceLidTightener}{left/right}

\providecommand{\ViveFaceUpperLidRaiser}{Eye\_*\_Wide}
\providecommand{\PicoFaceUpperLidRaiser}{EyeWide\_*}
\providecommand{\MetaFaceUpperLidRaiser}{UpperLidRaiser*}
\providecommand{\CommentFaceUpperLidRaiser}{left/right}

\providecommand{\ViveFaceLipCornerDepressor}{Mouth\_Sad\_*}
\providecommand{\PicoFaceLipCornerDepressor}{MouthFrown\_*}
\providecommand{\MetaFaceLipCornerDepressor}{LipCornerDepressor*}
\providecommand{\CommentFaceLipCornerDepressor}{left/right}

\providecommand{\ViveFaceLowerLipDepressor}{Mouth\_Lower\_Down*}
\providecommand{\PicoFaceLowerLipDepressor}{MouthLowerDown\_*}
\providecommand{\MetaFaceLowerLipDepressor}{LowerLipDepressor*}
\providecommand{\CommentFaceLowerLipDepressor}{left/right}

\providecommand{\ViveFaceUpperLipRaiser}{Mouth\_Upper\_Up*}
\providecommand{\PicoFaceUpperLipRaiser}{MouthUpperUp\_*}
\providecommand{\MetaFaceUpperLipRaiser}{UpperLipRaiser*}
\providecommand{\CommentFaceUpperLipRaiser}{left/right}

\providecommand{\ViveFaceLipCornerPuller}{Mouth\_Smile\_*}
\providecommand{\PicoFaceLipCornerPuller}{MouthSmile\_*}
\providecommand{\MetaFaceLipCornerPuller}{LipCornerPuller*}
\providecommand{\CommentFaceLipCornerPuller}{left/right}

\providecommand{\ViveFaceLipPucker}{Mouth\_Pout}
\providecommand{\PicoFaceLipPucker}{MouthPucker}
\providecommand{\MetaFaceLipPucker}{LipPucker*}
\providecommand{\CommentFaceLipPucker}{left/right}

\providecommand{\ViveFaceLipSuckB}{Mouth\_Lower\_Inside}
\providecommand{\PicoFaceLipSuckB}{MouthRollLower}
\providecommand{\MetaFaceLipSuckB}{LipSuck*B}
\providecommand{\CommentFaceLipSuckB}{left/right}

\providecommand{\ViveFaceLipSuckT}{Mouth\_Upper\_Inside}
\providecommand{\PicoFaceLipSuckT}{MouthRollUpper}
\providecommand{\MetaFaceLipSuckT}{LipSuck*T}
\providecommand{\CommentFaceLipSuckT}{left/right}

\providecommand{\ViveFaceMouth}{Mouth\_*\_*}
\providecommand{\PicoFaceMouth}{Mouth*}
\providecommand{\MetaFaceMouth}{Mouth*}
\providecommand{\CommentFaceMouth}{lower/upper}

\providecommand{\ViveFaceTongueOut}{Tongue\_LongStep*}
\providecommand{\PicoFaceTongueOut}{TongueOut}
\providecommand{\MetaFaceTongueOut}{TongueOut}
\providecommand{\CommentFaceTongueOut}{1, 2}

\providecommand{\ViveEyeLookDirection}{Gaze\_Direction\_*}
\providecommand{\PicoEyeLookDirection}{LookDirection*}
\providecommand{\MetaEyeLookDirection}{LookDirection*}
\providecommand{\CommentEyeLookDirection}{X,Y,Z; left, right}

\providecommand{\ViveEyePosition}{Gaze\_Origin\_MM\_*}
\providecommand{\PicoEyePosition}{Position*}
\providecommand{\MetaEyePosition}{Position*}
\providecommand{\CommentEyePosition}{X,Y,Z; left, right}

\providecommand{\ViveHeadDevicePosition}{DevicePosition*}
\providecommand{\PicoHeadDevicePosition}{DevicePosition*}
\providecommand{\MetaHeadDevicePosition}{DevicePosition*}
\providecommand{\CommentHeadDevicePosition}{X,Y,Z; left, right}

\providecommand{\ViveHeadDeviceRotation}{DeviceRotation*}
\providecommand{\PicoHeadDeviceRotation}{DeviceRotation*}
\providecommand{\MetaHeadDeviceRotation}{DeviceRotation*}
\providecommand{\CommentHeadDeviceRotation}{X,Y,Z,W; left, right}

\begin{table*}
\caption{Mapping from the device-dependent motion data attributes to the unified data format. For n-to-1 mappings from the devices to the unified format we use the mean of the directions.}\label{tab:unified_mapping}
\centering
\begin{tblr}{
  vlines,
  hline{1-2,17,19, 21} = {-}{},
  hline{3-22,23-24, 25-26} = {2-6}{},
}
\textbf{Type}    & \textbf{Unified}               & \textbf{Vive}  & \textbf{Pico}                          & \textbf{Meta}                          & \textbf{Direction (*)}\\
Facial  & CheekPuff             & \ViveFaceCheekPuff            & \PicoFaceCheekPuff            & \MetaFaceCheekPuff            & \CommentFaceCheekPuff             \\
        & EyeClosed*            & \ViveFaceEyeClosed            & \PicoFaceEyeClosed            & \MetaFaceEyeClosed            & \CommentFaceEyeClosed             \\
        & EyeLook*              & \ViveFaceEyeLook              & \PicoFaceEyeLook              & \MetaFaceEyeLook              & \CommentFaceEyeLook               \\
        & Jaw                   & \ViveFaceJaw                  & \PicoFaceJaw                  & \MetaFaceJaw                  & \CommentFaceJaw                   \\
        & LidTightener*         & \ViveFaceLidTightener         & \PicoFaceLidTightener         & \MetaFaceLidTightener         & \CommentFaceLidTightener          \\
        & UpperLidRaiser*       & \ViveFaceUpperLidRaiser       & \PicoFaceUpperLidRaiser       & \MetaFaceUpperLidRaiser       & \CommentFaceUpperLidRaiser        \\
        & LipCornerDepressor*   & \ViveFaceLipCornerDepressor   & \PicoFaceLipCornerDepressor   & \MetaFaceLipCornerDepressor   & \CommentFaceLipCornerDepressor    \\
        & LowerLipDepressor*    & \ViveFaceLowerLipDepressor    & \PicoFaceLowerLipDepressor    & \MetaFaceLowerLipDepressor    & \CommentFaceLowerLipDepressor     \\
        & UpperLipRaiser*       & \ViveFaceUpperLipRaiser       & \PicoFaceUpperLipRaiser       & \MetaFaceUpperLipRaiser       & \CommentFaceUpperLipRaiser        \\
        & LipCornerPuller*      & \ViveFaceLipCornerPuller      & \PicoFaceLipCornerPuller      & \MetaFaceLipCornerPuller      & \CommentFaceLipCornerPuller       \\
        & LipPucker*            & \ViveFaceLipPucker            & \PicoFaceLipPucker            & \MetaFaceLipPucker            & \CommentFaceLipPucker             \\
        & LipSuckB              & \ViveFaceLipSuckB             & \PicoFaceLipSuckB             & \MetaFaceLipSuckB             & \CommentFaceLipSuckB              \\
        & LipSuckT              & \ViveFaceLipSuckT             & \PicoFaceLipSuckT             & \MetaFaceLipSuckT             & \CommentFaceLipSuckT              \\
        & Mouth*                & \ViveFaceMouth                & \PicoFaceMouth                & \MetaFaceMouth                & \CommentFaceMouth                 \\
        & TongueOut             & \ViveFaceTongueOut            & \PicoFaceMouth                & \MetaFaceMouth                & \CommentFaceMouth                 \\
Eye     & LookDirection*        & \ViveEyeLookDirection         & \PicoEyeLookDirection         & \MetaEyeLookDirection         & \CommentEyeLookDirection          \\
        & Position*             & \ViveEyePosition              & \PicoEyePosition              & \MetaEyePosition              & \CommentEyePosition               \\
Head    & DevicePosition*       & \ViveHeadDevicePosition       & \PicoHeadDevicePosition       & \MetaHeadDevicePosition       & \CommentHeadDevicePosition        \\
        & DeviceRotation*       & \ViveHeadDeviceRotation       & \PicoHeadDeviceRotation       & \MetaHeadDeviceRotation       & \CommentHeadDeviceRotation        \\
\end{tblr}
\end{table*}

Upon the completion of each participant's session, our Unity project generated a unique directory containing the relevant data and metadata gathered during it.
This included the unsegmented face, eye, and head motion data, as well as the execution order and timestamp range of each task repetition, a microphone recording along with its metadata, and a log file.

Since the facial and eye motion data formats exported by the MR devices are not exactly the same, a unification step was necessary.
See Table~\ref{tab:unified_mapping} for the exact mapping for each MR headset.
For n-to-1 mappings from the devices to the unified format, we use the mean of the directions.
One example of this is the CheekPuff blendshape. The HTC Vive and Meta Quest Pro support CheekPuff for both sides of the face, while the Pico 4 Enterprise only returns one CheekPuff blendshape.

As our study consisted of tasks, we partitioned the unsegmented data of each participant into individual task-level segments.
Moreover, we segmented the aforementioned task-level segments which belonged to text tasks further into word- and phoneme-level segments.

\textbf{Task-Level Segmentation:}
First, the data was segmented by task.
To achieve this, we used the timestamp ranges stored during each task repetition.
When a participant started a task, a timestamp was saved to mark the start of execution. Then, when the participant finished the task, a second timestamp was saved to mark the end.
Since we stored the timestamp of when each sample of motion data was collected, we could identify which samples belonged to which task repetition in each motion data file.

\textbf{Text-Level Segmentation:}

We further segmented the verbal tasks into words (nursery rhymes only) and phonemes. To accomplish this, we aligned the speech recordings collected during task execution with the transcript of the performed task. We used a force alignment model to automatically perform this process on all verbal tasks and obtain the offset times for each word and phoneme uttered by the participant.

Due to synchronization problems between the audio recording and the recorded motion data, we first create a transcript of the entire recording by using WhisperX~\cite{bain2022whisperx}, an \ac{ASR} model, instead of aligning the recordings exclusively with the text of the verbal tasks. 
Another benefit of this approach is that we can also account for unforeseeable words that were possibly uttered at the beginning of the recording, and for which we did not have a transcript before. Then, we locate the verbal tasks in the transcript and correct any errors using the text of the specific task.

These transcriptions were then used as input for the \ac{MFA}~\cite{mcauliffe17_interspeech}, along with the full recordings. By being given the full transcriptions, the model accurately aligned them to the audio recordings and returned the offsets of when each word and phoneme was uttered.

As a last step, we had to convert the alignment offsets in the audio recordings to the actual timestamp ranges in the motion data files. To do this effectively, we interpolated the start and stop timestamps of the text tasks in the data with the start and stop alignment offsets of the same text tasks obtained from \ac{MFA}. As a result, we could segment the text task data into word and phoneme segments.

\subsection{Data Availability}

\providecommand{\TTotal}{x}
\providecommand{\TGroupA}{x}
\providecommand{\TGroupB}{y}
\providecommand{\TDeviceV}{x}
\providecommand{\TDeviceP}{y}
\providecommand{\TDeviceM}{z}
\providecommand{\TID}{}
\providecommand{\TFstSesh}{x}
\providecommand{\TSndSesh}{y}
\providecommand{\TTrdSesh}{z}

\providecommand{\WTotal}{x}
\providecommand{\WGroupA}{x}
\providecommand{\WGroupB}{y}
\providecommand{\WDeviceV}{x}
\providecommand{\WDeviceP}{y}
\providecommand{\WDeviceM}{z}
\providecommand{\WID}{y}
\providecommand{\WFstSesh}{x}
\providecommand{\WSndSesh}{y}
\providecommand{\WTrdSesh}{z}

\providecommand{\PhTotal}{x}
\providecommand{\PhGroupA}{x}
\providecommand{\PhGroupB}{y}
\providecommand{\PhDeviceV}{x}
\providecommand{\PhDeviceP}{y}
\providecommand{\PhDeviceM}{z}
\providecommand{\PhID}{y}
\providecommand{\PhFstSesh}{x}
\providecommand{\PhSndSesh}{y}
\providecommand{\PhTrdSesh}{z}

\begin{table*}
\caption{Overview of the dataset regarding the amount of samples it comprises.}\label{tab:dataset}
\centering
\begin{tblr}{
  colspec={Q[c] Q[c] Q[c] Q[c] Q[c] Q[c] Q[c] Q[c] Q[c] Q[c]},
  vlines,
  hlines
}
\SetCell[r=2]{c} \textbf{Segmentation}  & \SetCell[r=2]{c} \textbf{Total}    & \SetCell[c=2]{c} \textbf{Per Group} & &   \SetCell[c=3]{c} \textbf{Per Device}  & & &  \SetCell[c=3]{c} \textbf{Per Session} & &                     \\
                        &                           & \textbf{A}             & \textbf{B}                & \textbf{Vive}          & \textbf{Pico}             &  \textbf{Meta}             &    \textbf{0}             & \textbf{1}             & \textbf{2}             \\
Recordings    & 499                       & 232      & 267                & 132       & 127            & 240  & 229   & 150      & 120                          \\
Tasks                   & 19296                   & 8883     & 10413         & 5175  & 4905    & 9216     & 8136     & 6750 & 4410  \\
Words                   & 197255  & 88477 & 108778 & 45087  & 51631  & 100537 & 82814 & 67362  &  47079 \\   
\end{tblr}
\end{table*}

In total, we recorded 259 sessions. 19 of these sessions were missing one headset recording, resulting in a total of 499 individual headset recordings. Table~\ref{tab:dataset} provides an overview of the number of samples segmented as described above. The dataset will be published alongside this paper.

\section{Evaluation}\label{sec:evaluation}

Here, we present the evaluation that we performed on the dataset. Our main goal is to investigate the types of privacy inferences that can be made from facial motion data. However, we also perform the same experiment on eye gaze and head motion data to allow for comparison. 
First, we present the experiments we performed. 
Next, we detail the methodology for the biometric recognition system. Lastly, we present the results of the experiments.

\subsection{Experiments}
We first want to learn if identification from facial motion data collected with \ac{MR} headsets is possible.
The prior work on facial motion videos (see Section~\ref{sec:facial_identification}) and on eye gaze identification (see Section~\ref{sec:eye_gaze_identification}) suggests that this should be possible.

In our first Experiment \textbf{E1}, we wish to learn if individuals can be identified from their facial motion data. For this experiment, we will investigate the identification for each headset separately, as well as all headsets together. We then investigate the influence of the sessions for Experiment \textbf{E2}. Hence, we use the first two sessions for training the biometric recognition system and then only test on the third session. This way we can see if the identification is stable over time. Next, we examine in Experiment \textbf{E3} if we can re-identify individuals when they start using different headsets, giving us insights how dependent the identification is on the headset type and if it can be generalized across \ac{MR} headset types.

Besides identification, the related work (see Section~\ref{sec:facial_exp_recog}) suggests that it should be possible to infer the facial expression and therefore we expect that it is possible to infer the emotion displayed in the non-verbal tasks.

In Experiment \textbf{E4}, we test how good we can recognize the emotions displayed in our non-verbal tasks. Further, we also test if we can correctly classify which verbal task was performed. In Experiment \textbf{E5}, we investigate if the \ac{MR} headset type can be inferred from the data collected. All headset data has the same format due to the unified data format, however, we expect that it is easy to infer which headset is being used due to device specific quirks. Then, we look at the inference of sensitive attributes about the user of the \ac{MR} headset in Experiment \textbf{E6}. Here, we seek to infer the sex, English level, and personality trait of the user.

Lastly, we perform two experiments to better understand the identification from facial motion data. In Experiment \textbf{E7}, we perform the identification only on the verbal tasks or only on the non-verbal tasks to see which task type works better for identification. And in Experiment \textbf{E8}, we test how good we can identify individuals when we combine the facial motion data with the eye gaze and head motion data.

\subsection{Data Preparation \& Splitting}

For our evaluation, we use task-level segmentation of our dataset in the unified data format. We filter out the recordings performed on January 22, 2025, as some of the Vive's facial motion data values are missing. We then remove the timestamp column from the remaining samples and resample each one to 100 frames, normalizing the size of all samples.

Next, we split the data into training and testing datasets for the biometric recognition model. The testing dataset is used exclusively to calculate the model's final performance. Since different experiments require different data splits, we use multiple splits:

\textbf{Random:} For the random split type, we randomly split all samples, allocating 80\% to the training dataset and 20\% to the testing dataset.

\textbf{Session:} For the sessions split-type, we use the recordings from the first two sessions from each participant as the training dataset and the last session as the testing dataset.

\textbf{Leave-one-headset-out-per-participant (LHPP):} The LHPP split type uses two \ac{MR} headsets per participant for the training dataset and one MR headset for the testing dataset. This allows the biometric recognition model to learn to recognize specific participants and to use data from each \ac{MR} headset type.

\textbf{Participant:} The participant split type allocates 80\% of participants to the training dataset and 20\% to the test dataset. This type of split is used for attribute inference experiments to prevent cross-contamination of the results, e.g. the model learning to recognize attributes by identifying the specific participant.

\subsection{Biometric Recognition Models}

For our experiments, we use three different machine learning models as a biometric recognition system. 
The first is a \textbf{simple} fully connected neural network that receives each sample as a single vector. 
This neural network consists of at least two fully connected linear layers and a variable number of hidden layers, which are determined via hyper parameter optimization. 
After each linear layer, we use a \ac{ReLu} activation function, as well as a dropout layer, to prevent overfitting.
The second model is a \textit{Long Short-Term Memory} (\textbf{LSTM}) that processes each sample frame-by-frame. To determine the most likely class, we first use a linear layer to reduce the size of the output vector to the number of classes.
The third model is \textbf{EYKT}~\cite{9865991}, a DenseNet-based architecture. Between each convolution block, the network uses batch normalization and the \ac{ReLu} activation function.
All networks use log softmax to perform the final classification step.

The training dataset is randomly split into a main training dataset and a validation dataset for model training. The main training dataset contains 90\% of the data, and the validation dataset contains 10\%. Each model is trained for a maximum of 100 epochs with early stopping if the validation accuracy does not increase for 10 epochs. We use negative log likelihood loss as the loss function and 1280 samples as the batch size.

We determine the best model parameters for each experiment by performing parameter optimization for 100 steps. See Table~\ref{tab:para_opt} for the optimized parameters. After optimization, we use the model with the best performance on the validation dataset and run it on the testing dataset to determine the final accuracy for each experiment.

\begin{table}[h]
\centering
\caption{Overview of the optimized parameters}\label{tab:para_opt}
\begin{tabular}{|l|l|l|}
\hline
\textbf{Parameter}      & \textbf{Range} & \textbf{Note} \\ \hline
Layer Size              & 10-256         & Only Simple \& LSTM\\ \hline
Hidden Layers           & 0-2            & Only Simple \& LSTM\\ \hline
Learning Rate Step Size & 10-100         & All \\ \hline
Learning Rate Alpha     & 0.01-1         & All \\ \hline
Optimizer Learning Rate & 0.0001-0.1     & All \\ \hline
Weight Decay            & 0.00001-0.01   & All \\ \hline
\end{tabular}
\end{table}

\subsection{Implementation}

We implemented the biometric recognition models using Python (3.12) and PyTorch (2.6.0). As learning optimizer, we used Adam, and for the parameter optimization we used Optuna (4.3). The code used for our evaluation will be published alongside this paper.

\subsection{Results}\label{sec:results}

Here, we present our evaluation results. As a metric, we always use the accuracy, which is defined as the correct classifications divided by all classifications. Further, we also always give the percentage of the largest class in the experiment-specific data split as the chance level.

\providecommand{\BaselineFaceSimpleVive}{$0.78$}
\definecolor{cBaselineFaceSimpleVive}{rgb}{0.430983,0.808473,0.346476}
\providecommand{\BaselineFaceSimplePico}{$0.83$}
\definecolor{cBaselineFaceSimplePico}{rgb}{0.555484,0.840254,0.269281}
\providecommand{\BaselineFaceSimpleMeta}{$0.63$}
\definecolor{cBaselineFaceSimpleMeta}{rgb}{0.162016,0.687316,0.499129}
\providecommand{\BaselineFaceSimpleAll}{$0.68$}
\definecolor{cBaselineFaceSimpleAll}{rgb}{0.232815,0.732247,0.459277}
\providecommand{\BaselineFaceSimpleChance}{$0.02$}
\providecommand{\BaselineFaceLSTMVive}{$0.69$}
\definecolor{cBaselineFaceLSTMVive}{rgb}{0.24607,0.73891,0.452024}
\providecommand{\BaselineFaceLSTMPico}{$0.77$}
\definecolor{cBaselineFaceLSTMPico}{rgb}{0.412913,0.803041,0.357269}
\providecommand{\BaselineFaceLSTMMeta}{$0.59$}
\definecolor{cBaselineFaceLSTMMeta}{rgb}{0.130067,0.651384,0.521608}
\providecommand{\BaselineFaceLSTMAll}{$0.58$}
\definecolor{cBaselineFaceLSTMAll}{rgb}{0.12478,0.640461,0.527068}
\providecommand{\BaselineFaceLSTMChance}{$0.02$}
\providecommand{\BaselineFaceEYKTVive}{$0.94$}
\definecolor{cBaselineFaceEYKTVive}{rgb}{0.845561,0.887322,0.099702}
\providecommand{\BaselineFaceEYKTPico}{$0.98$}
\definecolor{cBaselineFaceEYKTPico}{rgb}{0.945636,0.899815,0.112838}
\providecommand{\BaselineFaceEYKTMeta}{$0.88$}
\definecolor{cBaselineFaceEYKTMeta}{rgb}{0.688944,0.865448,0.182725}
\providecommand{\BaselineFaceEYKTAll}{$0.9$}
\definecolor{cBaselineFaceEYKTAll}{rgb}{0.741388,0.873449,0.149561}
\providecommand{\BaselineFaceEYKTChance}{$0.02$}
\providecommand{\BaselineHeadSimpleVive}{$0.99$}
\definecolor{cBaselineHeadSimpleVive}{rgb}{0.974417,0.90359,0.130215}
\providecommand{\BaselineHeadSimplePico}{$0.88$}
\definecolor{cBaselineHeadSimplePico}{rgb}{0.688944,0.865448,0.182725}
\providecommand{\BaselineHeadSimpleMeta}{$0.76$}
\definecolor{cBaselineHeadSimpleMeta}{rgb}{0.386433,0.794644,0.372886}
\providecommand{\BaselineHeadSimpleAll}{$0.7$}
\definecolor{cBaselineHeadSimpleAll}{rgb}{0.266941,0.748751,0.440573}
\providecommand{\BaselineHeadSimpleChance}{$0.02$}
\providecommand{\BaselineHeadLSTMVive}{$0.94$}
\definecolor{cBaselineHeadLSTMVive}{rgb}{0.845561,0.887322,0.099702}
\providecommand{\BaselineHeadLSTMPico}{$0.78$}
\definecolor{cBaselineHeadLSTMPico}{rgb}{0.430983,0.808473,0.346476}
\providecommand{\BaselineHeadLSTMMeta}{$0.76$}
\definecolor{cBaselineHeadLSTMMeta}{rgb}{0.386433,0.794644,0.372886}
\providecommand{\BaselineHeadLSTMAll}{$0.8$}
\definecolor{cBaselineHeadLSTMAll}{rgb}{0.477504,0.821444,0.318195}
\providecommand{\BaselineHeadLSTMChance}{$0.02$}
\providecommand{\BaselineHeadEYKTVive}{$1.0$}
\definecolor{cBaselineHeadEYKTVive}{rgb}{0.993248,0.906157,0.143936}
\providecommand{\BaselineHeadEYKTPico}{$0.95$}
\definecolor{cBaselineHeadEYKTPico}{rgb}{0.876168,0.891125,0.09525}
\providecommand{\BaselineHeadEYKTMeta}{$0.98$}
\definecolor{cBaselineHeadEYKTMeta}{rgb}{0.945636,0.899815,0.112838}
\providecommand{\BaselineHeadEYKTAll}{$0.95$}
\definecolor{cBaselineHeadEYKTAll}{rgb}{0.876168,0.891125,0.09525}
\providecommand{\BaselineHeadEYKTChance}{$0.02$}
\providecommand{\BaselineEyeSimpleVive}{$0.87$}
\definecolor{cBaselineEyeSimpleVive}{rgb}{0.657642,0.860219,0.203082}
\providecommand{\BaselineEyeSimplePico}{$0.7$}
\definecolor{cBaselineEyeSimplePico}{rgb}{0.266941,0.748751,0.440573}
\providecommand{\BaselineEyeSimpleMeta}{$0.45$}
\definecolor{cBaselineEyeSimpleMeta}{rgb}{0.144759,0.519093,0.556572}
\providecommand{\BaselineEyeSimpleAll}{$0.54$}
\definecolor{cBaselineEyeSimpleAll}{rgb}{0.119738,0.603785,0.5414}
\providecommand{\BaselineEyeSimpleChance}{$0.02$}
\providecommand{\BaselineEyeLSTMVive}{$0.86$}
\definecolor{cBaselineEyeLSTMVive}{rgb}{0.636902,0.856542,0.21662}
\providecommand{\BaselineEyeLSTMPico}{$0.59$}
\definecolor{cBaselineEyeLSTMPico}{rgb}{0.130067,0.651384,0.521608}
\providecommand{\BaselineEyeLSTMMeta}{$0.47$}
\definecolor{cBaselineEyeLSTMMeta}{rgb}{0.13777,0.537492,0.554906}
\providecommand{\BaselineEyeLSTMAll}{$0.49$}
\definecolor{cBaselineEyeLSTMAll}{rgb}{0.131172,0.555899,0.552459}
\providecommand{\BaselineEyeLSTMChance}{$0.02$}
\providecommand{\BaselineEyeEYKTVive}{$1.0$}
\definecolor{cBaselineEyeEYKTVive}{rgb}{0.993248,0.906157,0.143936}
\providecommand{\BaselineEyeEYKTPico}{$0.87$}
\definecolor{cBaselineEyeEYKTPico}{rgb}{0.657642,0.860219,0.203082}
\providecommand{\BaselineEyeEYKTMeta}{$0.92$}
\definecolor{cBaselineEyeEYKTMeta}{rgb}{0.79376,0.880678,0.120005}
\providecommand{\BaselineEyeEYKTAll}{$0.78$}
\definecolor{cBaselineEyeEYKTAll}{rgb}{0.430983,0.808473,0.346476}
\providecommand{\BaselineEyeEYKTChance}{$0.02$}

\begin{table}
\caption{Identification accuracy using a random split}\label{tab:Baseline}
\centering
\begin{tblr}{
  cell{2}{1} = {r=3}{},
  cell{2}{3} = {cBaselineFaceSimpleVive},
  cell{2}{4} = {cBaselineFaceSimplePico},
  cell{2}{5} = {cBaselineFaceSimpleMeta},
  cell{2}{6} = {cBaselineFaceSimpleAll},
  cell{3}{3} = {cBaselineFaceLSTMVive},
  cell{3}{4} = {cBaselineFaceLSTMPico},
  cell{3}{5} = {cBaselineFaceLSTMMeta},
  cell{3}{6} = {cBaselineFaceLSTMAll},
  cell{4}{3} = {cBaselineFaceEYKTVive},
  cell{4}{4} = {cBaselineFaceEYKTPico},
  cell{4}{5} = {cBaselineFaceEYKTMeta},
  cell{4}{6} = {cBaselineFaceEYKTAll},
  cell{5}{1} = {r=3}{},
  cell{5}{3} = {cBaselineEyeSimpleVive},
  cell{5}{4} = {cBaselineEyeSimplePico},
  cell{5}{5} = {cBaselineEyeSimpleMeta},
  cell{5}{6} = {cBaselineEyeSimpleAll},
  cell{6}{3} = {cBaselineEyeLSTMVive},
  cell{6}{4} = {cBaselineEyeLSTMPico},
  cell{6}{5} = {cBaselineEyeLSTMMeta},
  cell{6}{6} = {cBaselineEyeLSTMAll},
  cell{7}{3} = {cBaselineEyeEYKTVive},
  cell{7}{4} = {cBaselineEyeEYKTPico},
  cell{7}{5} = {cBaselineEyeEYKTMeta},
  cell{7}{6} = {cBaselineEyeEYKTAll},
  cell{8}{1} = {r=3}{},
  cell{8}{3} = {cBaselineHeadSimpleVive},
  cell{8}{4} = {cBaselineHeadSimplePico},
  cell{8}{5} = {cBaselineHeadSimpleMeta},
  cell{8}{6} = {cBaselineHeadSimpleAll},
  cell{9}{3} = {cBaselineHeadLSTMVive},
  cell{9}{4} = {cBaselineHeadLSTMPico},
  cell{9}{5} = {cBaselineHeadLSTMMeta},
  cell{9}{6} = {cBaselineHeadLSTMAll},
  cell{10}{3} = {cBaselineHeadEYKTVive},
  cell{10}{4} = {cBaselineHeadEYKTPico},
  cell{10}{5} = {cBaselineHeadEYKTMeta},
  cell{10}{6} = {cBaselineHeadEYKTAll},
  vlines,
  hline{1-2,5,8,11} = {-}{},
  hline{3-4,6-7,9-10} = {2-7}{},
}
Data Type & Model  & Vive & Pico & Meta & All & Chance \\
Facial      & Simple & \BaselineFaceSimpleVive  &  \BaselineFaceSimplePico & \BaselineFaceSimpleMeta &  \BaselineFaceSimpleAll & \BaselineFaceSimpleChance \\
          & LSTM   & \BaselineFaceLSTMVive  &  \BaselineFaceLSTMPico & \BaselineFaceLSTMMeta &  \BaselineFaceLSTMAll & \BaselineFaceLSTMChance \\
          & EYKT   & \BaselineFaceEYKTVive  &  \BaselineFaceEYKTPico & \BaselineFaceEYKTMeta &  \BaselineFaceEYKTAll & \BaselineFaceEYKTChance \\
Eye       & Simple & \BaselineEyeSimpleVive  &  \BaselineEyeSimplePico & \BaselineEyeSimpleMeta &  \BaselineEyeSimpleAll & \BaselineEyeSimpleChance \\
          & LSTM   & \BaselineEyeLSTMVive  &  \BaselineEyeLSTMPico & \BaselineEyeLSTMMeta &  \BaselineEyeLSTMAll & \BaselineEyeLSTMChance \\
          & EYKT   & \BaselineEyeEYKTVive  &  \BaselineEyeEYKTPico & \BaselineEyeEYKTMeta &  \BaselineEyeEYKTAll & \BaselineEyeEYKTChance \\
Head      & Simple & \BaselineHeadSimpleVive  &  \BaselineHeadSimplePico & \BaselineHeadSimpleMeta &  \BaselineHeadSimpleAll & \BaselineHeadSimpleChance \\
          & LSTM   & \BaselineHeadLSTMVive  &  \BaselineHeadLSTMPico & \BaselineHeadLSTMMeta &  \BaselineHeadLSTMAll & \BaselineHeadLSTMChance \\
          & EYKT   & \BaselineHeadEYKTVive  &  \BaselineHeadEYKTPico & \BaselineHeadEYKTMeta &  \BaselineHeadEYKTAll & \BaselineHeadEYKTChance 
\end{tblr}
\end{table}

For our Experiment \textbf{E1} (see Table~\ref{tab:Baseline}), we used the random split to gain general understanding of how well the identification works. For the facial data, we find that the Pico achieves the highest identification of 98\%, the Vive achieves 94\%, the Meta achieves 88\%, and using all headsets together we achieve 90\%. This fulfills our expectation that identification on facial motion data is possible.
Comparing to the eye and head data types, we find that for both we achieve 100\% identification for the Vive and the EYKT model. In general, we can observe that the identification works for all data types, and all headsets, with the head motion data performing the best in general, though the facial and eye motions are not far behind.

\providecommand{\SessionFaceSimpleVive}{\textcolor{white}{$0.11$}}
\definecolor{cSessionFaceSimpleVive}{rgb}{0.281412,0.155834,0.469201}
\providecommand{\SessionFaceSimplePico}{\textcolor{white}{$0.26$}}
\definecolor{cSessionFaceSimplePico}{rgb}{0.225863,0.330805,0.547314}
\providecommand{\SessionFaceSimpleMeta}{\textcolor{white}{$0.29$}}
\definecolor{cSessionFaceSimpleMeta}{rgb}{0.210503,0.363727,0.552206}
\providecommand{\SessionFaceSimpleAll}{\textcolor{white}{$0.2$}}
\definecolor{cSessionFaceSimpleAll}{rgb}{0.253935,0.265254,0.529983}
\providecommand{\SessionFaceSimpleChance}{$0.04$}
\providecommand{\SessionFaceLSTMVive}{\textcolor{white}{$0.11$}}
\definecolor{cSessionFaceLSTMVive}{rgb}{0.281412,0.155834,0.469201}
\providecommand{\SessionFaceLSTMPico}{\textcolor{white}{$0.11$}}
\definecolor{cSessionFaceLSTMPico}{rgb}{0.281412,0.155834,0.469201}
\providecommand{\SessionFaceLSTMMeta}{\textcolor{white}{$0.24$}}
\definecolor{cSessionFaceLSTMMeta}{rgb}{0.235526,0.309527,0.542944}
\providecommand{\SessionFaceLSTMAll}{\textcolor{white}{$0.17$}}
\definecolor{cSessionFaceLSTMAll}{rgb}{0.26658,0.228262,0.514349}
\providecommand{\SessionFaceLSTMChance}{$0.04$}
\providecommand{\SessionFaceEYKTVive}{\textcolor{white}{$0.14$}}
\definecolor{cSessionFaceEYKTVive}{rgb}{0.276194,0.190074,0.493001}
\providecommand{\SessionFaceEYKTPico}{\textcolor{white}{$0.23$}}
\definecolor{cSessionFaceEYKTPico}{rgb}{0.241237,0.296485,0.539709}
\providecommand{\SessionFaceEYKTMeta}{$0.43$}
\definecolor{cSessionFaceEYKTMeta}{rgb}{0.151918,0.500685,0.557587}
\providecommand{\SessionFaceEYKTAll}{\textcolor{white}{$0.27$}}
\definecolor{cSessionFaceEYKTAll}{rgb}{0.220057,0.343307,0.549413}
\providecommand{\SessionFaceEYKTChance}{$0.04$}
\providecommand{\SessionHeadSimpleVive}{\textcolor{white}{$0.0$}}
\definecolor{cSessionHeadSimpleVive}{rgb}{0.267004,0.004874,0.329415}
\providecommand{\SessionHeadSimplePico}{\textcolor{white}{$0.02$}}
\definecolor{cSessionHeadSimplePico}{rgb}{0.273809,0.031497,0.358853}
\providecommand{\SessionHeadSimpleMeta}{\textcolor{white}{$0.16$}}
\definecolor{cSessionHeadSimpleMeta}{rgb}{0.270595,0.214069,0.507052}
\providecommand{\SessionHeadSimpleAll}{\textcolor{white}{$0.08$}}
\definecolor{cSessionHeadSimpleAll}{rgb}{0.283197,0.11568,0.436115}
\providecommand{\SessionHeadSimpleChance}{$0.04$}
\providecommand{\SessionHeadLSTMVive}{\textcolor{white}{$0.0$}}
\definecolor{cSessionHeadLSTMVive}{rgb}{0.267004,0.004874,0.329415}
\providecommand{\SessionHeadLSTMPico}{\textcolor{white}{$0.01$}}
\definecolor{cSessionHeadLSTMPico}{rgb}{0.269944,0.014625,0.341379}
\providecommand{\SessionHeadLSTMMeta}{\textcolor{white}{$0.15$}}
\definecolor{cSessionHeadLSTMMeta}{rgb}{0.273006,0.20452,0.501721}
\providecommand{\SessionHeadLSTMAll}{\textcolor{white}{$0.06$}}
\definecolor{cSessionHeadLSTMAll}{rgb}{0.281924,0.089666,0.412415}
\providecommand{\SessionHeadLSTMChance}{$0.04$}
\providecommand{\SessionHeadEYKTVive}{\textcolor{white}{$0.0$}}
\definecolor{cSessionHeadEYKTVive}{rgb}{0.267004,0.004874,0.329415}
\providecommand{\SessionHeadEYKTPico}{\textcolor{white}{$0.01$}}
\definecolor{cSessionHeadEYKTPico}{rgb}{0.269944,0.014625,0.341379}
\providecommand{\SessionHeadEYKTMeta}{\textcolor{white}{$0.16$}}
\definecolor{cSessionHeadEYKTMeta}{rgb}{0.270595,0.214069,0.507052}
\providecommand{\SessionHeadEYKTAll}{\textcolor{white}{$0.07$}}
\definecolor{cSessionHeadEYKTAll}{rgb}{0.282656,0.100196,0.42216}
\providecommand{\SessionHeadEYKTChance}{$0.04$}
\providecommand{\SessionEyeSimpleVive}{\textcolor{white}{$0.14$}}
\definecolor{cSessionEyeSimpleVive}{rgb}{0.276194,0.190074,0.493001}
\providecommand{\SessionEyeSimplePico}{\textcolor{white}{$0.03$}}
\definecolor{cSessionEyeSimplePico}{rgb}{0.276022,0.044167,0.370164}
\providecommand{\SessionEyeSimpleMeta}{\textcolor{white}{$0.03$}}
\definecolor{cSessionEyeSimpleMeta}{rgb}{0.276022,0.044167,0.370164}
\providecommand{\SessionEyeSimpleAll}{\textcolor{white}{$0.06$}}
\definecolor{cSessionEyeSimpleAll}{rgb}{0.281924,0.089666,0.412415}
\providecommand{\SessionEyeSimpleChance}{$0.04$}
\providecommand{\SessionEyeLSTMVive}{\textcolor{white}{$0.12$}}
\definecolor{cSessionEyeLSTMVive}{rgb}{0.280255,0.165693,0.476498}
\providecommand{\SessionEyeLSTMPico}{\textcolor{white}{$0.02$}}
\definecolor{cSessionEyeLSTMPico}{rgb}{0.273809,0.031497,0.358853}
\providecommand{\SessionEyeLSTMMeta}{\textcolor{white}{$0.07$}}
\definecolor{cSessionEyeLSTMMeta}{rgb}{0.282656,0.100196,0.42216}
\providecommand{\SessionEyeLSTMAll}{\textcolor{white}{$0.05$}}
\definecolor{cSessionEyeLSTMAll}{rgb}{0.280267,0.073417,0.397163}
\providecommand{\SessionEyeLSTMChance}{$0.04$}
\providecommand{\SessionEyeEYKTVive}{\textcolor{white}{$0.1$}}
\definecolor{cSessionEyeEYKTVive}{rgb}{0.282623,0.140926,0.457517}
\providecommand{\SessionEyeEYKTPico}{\textcolor{white}{$0.06$}}
\definecolor{cSessionEyeEYKTPico}{rgb}{0.281924,0.089666,0.412415}
\providecommand{\SessionEyeEYKTMeta}{\textcolor{white}{$0.1$}}
\definecolor{cSessionEyeEYKTMeta}{rgb}{0.282623,0.140926,0.457517}
\providecommand{\SessionEyeEYKTAll}{\textcolor{white}{$0.09$}}
\definecolor{cSessionEyeEYKTAll}{rgb}{0.283072,0.130895,0.449241}
\providecommand{\SessionEyeEYKTChance}{$0.04$}

\begin{table}
\caption{Identification accuracy using a session split}\label{tab:Session}
\centering
\begin{tblr}{
  cell{2}{1} = {r=3}{},
  cell{2}{3} = {cSessionFaceSimpleVive},
  cell{2}{4} = {cSessionFaceSimplePico},
  cell{2}{5} = {cSessionFaceSimpleMeta},
  cell{2}{6} = {cSessionFaceSimpleAll},
  cell{3}{3} = {cSessionFaceLSTMVive},
  cell{3}{4} = {cSessionFaceLSTMPico},
  cell{3}{5} = {cSessionFaceLSTMMeta},
  cell{3}{6} = {cSessionFaceLSTMAll},
  cell{4}{3} = {cSessionFaceEYKTVive},
  cell{4}{4} = {cSessionFaceEYKTPico},
  cell{4}{5} = {cSessionFaceEYKTMeta},
  cell{4}{6} = {cSessionFaceEYKTAll},
  cell{5}{1} = {r=3}{},
  cell{5}{3} = {cSessionEyeSimpleVive},
  cell{5}{4} = {cSessionEyeSimplePico},
  cell{5}{5} = {cSessionEyeSimpleMeta},
  cell{5}{6} = {cSessionEyeSimpleAll},
  cell{6}{3} = {cSessionEyeLSTMVive},
  cell{6}{4} = {cSessionEyeLSTMPico},
  cell{6}{5} = {cSessionEyeLSTMMeta},
  cell{6}{6} = {cSessionEyeLSTMAll},
  cell{7}{3} = {cSessionEyeEYKTVive},
  cell{7}{4} = {cSessionEyeEYKTPico},
  cell{7}{5} = {cSessionEyeEYKTMeta},
  cell{7}{6} = {cSessionEyeEYKTAll},
  cell{8}{1} = {r=3}{},
  cell{8}{3} = {cSessionHeadSimpleVive},
  cell{8}{4} = {cSessionHeadSimplePico},
  cell{8}{5} = {cSessionHeadSimpleMeta},
  cell{8}{6} = {cSessionHeadSimpleAll},
  cell{9}{3} = {cSessionHeadLSTMVive},
  cell{9}{4} = {cSessionHeadLSTMPico},
  cell{9}{5} = {cSessionHeadLSTMMeta},
  cell{9}{6} = {cSessionHeadLSTMAll},
  cell{10}{3} = {cSessionHeadEYKTVive},
  cell{10}{4} = {cSessionHeadEYKTPico},
  cell{10}{5} = {cSessionHeadEYKTMeta},
  cell{10}{6} = {cSessionHeadEYKTAll},
  vlines,
  hline{1-2,5,8,11} = {-}{},
  hline{3-4,6-7,9-10} = {2-7}{},
}
Data Type & Model  & Vive & Pico & Meta & All & Chance \\
Facial      & Simple & \SessionFaceSimpleVive  &  \SessionFaceSimplePico & \SessionFaceSimpleMeta &  \SessionFaceSimpleAll & \SessionFaceSimpleChance \\
          & LSTM   & \SessionFaceLSTMVive  &  \SessionFaceLSTMPico & \SessionFaceLSTMMeta &  \SessionFaceLSTMAll & \SessionFaceLSTMChance \\
          & EYKT   & \SessionFaceEYKTVive  &  \SessionFaceEYKTPico & \SessionFaceEYKTMeta &  \SessionFaceEYKTAll & \SessionFaceEYKTChance \\
Eye       & Simple & \SessionEyeSimpleVive  &  \SessionEyeSimplePico & \SessionEyeSimpleMeta &  \SessionEyeSimpleAll & \SessionEyeSimpleChance \\
          & LSTM   & \SessionEyeLSTMVive  &  \SessionEyeLSTMPico & \SessionEyeLSTMMeta &  \SessionEyeLSTMAll & \SessionEyeLSTMChance \\
          & EYKT   & \SessionEyeEYKTVive  &  \SessionEyeEYKTPico & \SessionEyeEYKTMeta &  \SessionEyeEYKTAll & \SessionEyeEYKTChance \\
Head      & Simple & \SessionHeadSimpleVive  &  \SessionHeadSimplePico & \SessionHeadSimpleMeta &  \SessionHeadSimpleAll & \SessionHeadSimpleChance \\
          & LSTM   & \SessionHeadLSTMVive  &  \SessionHeadLSTMPico & \SessionHeadLSTMMeta &  \SessionHeadLSTMAll & \SessionHeadLSTMChance \\
          & EYKT   & \SessionHeadEYKTVive  &  \SessionHeadEYKTPico & \SessionHeadEYKTMeta &  \SessionHeadEYKTAll & \SessionHeadEYKTChance 
\end{tblr}
\end{table}

Moving on to Experiment \textbf{E2} (see Table~\ref{tab:Session}), we now split the data according to their sessions into training and testing dataset. In general, we can see for the face data that all headsets and model combinations exceed the chance level for identification. The best result is 43\% balanced accuracy for the face data of the Meta when using the EYKT model. We conclude that the identification across sessions is possible, but most of the learned features from E1 identify the specific session and are not general for the individual. Comparing E1 and E2 results, it is also interesting to see that in E2 the Meta performs far better for facial motions, while in E1 it has the worst performance of all three headset types. %

\providecommand{\DeviceToDeviceFaceSimple}{$0.61$}
\definecolor{cDeviceToDeviceFaceSimple}{rgb}{0.143303,0.669459,0.511215}
\providecommand{\DeviceToDeviceFaceLSTM}{$0.52$}
\definecolor{cDeviceToDeviceFaceLSTM}{rgb}{0.122606,0.585371,0.546557}
\providecommand{\DeviceToDeviceFaceEYKT}{$0.63$}
\definecolor{cDeviceToDeviceFaceEYKT}{rgb}{0.162016,0.687316,0.499129}
\providecommand{\DeviceToDeviceFaceChance}{$0.02$}
\providecommand{\DeviceToDeviceHeadSimple}{$0.49$}
\definecolor{cDeviceToDeviceHeadSimple}{rgb}{0.131172,0.555899,0.552459}
\providecommand{\DeviceToDeviceHeadLSTM}{$0.46$}
\definecolor{cDeviceToDeviceHeadLSTM}{rgb}{0.141935,0.526453,0.555991}
\providecommand{\DeviceToDeviceHeadEYKT}{$0.63$}
\definecolor{cDeviceToDeviceHeadEYKT}{rgb}{0.162016,0.687316,0.499129}
\providecommand{\DeviceToDeviceHeadChance}{$0.02$}
\providecommand{\DeviceToDeviceEyeSimple}{$0.45$}
\definecolor{cDeviceToDeviceEyeSimple}{rgb}{0.144759,0.519093,0.556572}
\providecommand{\DeviceToDeviceEyeLSTM}{$0.47$}
\definecolor{cDeviceToDeviceEyeLSTM}{rgb}{0.13777,0.537492,0.554906}
\providecommand{\DeviceToDeviceEyeEYKT}{$0.65$}
\definecolor{cDeviceToDeviceEyeEYKT}{rgb}{0.185783,0.704891,0.485273}
\providecommand{\DeviceToDeviceEyeChance}{$0.02$}

\begin{table}
\caption{Identification accuracy using the LHPP split for all headsets}\label{tab:DeviceToDevice}
\centering
\begin{tblr}{
  cell{2}{2} = {cDeviceToDeviceFaceSimple},
  cell{2}{3} = {cDeviceToDeviceFaceLSTM},
  cell{2}{4} = {cDeviceToDeviceFaceEYKT},
  cell{3}{2} = {cDeviceToDeviceEyeSimple},
  cell{3}{3} = {cDeviceToDeviceEyeLSTM},
  cell{3}{4} = {cDeviceToDeviceEyeEYKT},
  cell{4}{2} = {cDeviceToDeviceHeadSimple},
  cell{4}{3} = {cDeviceToDeviceHeadLSTM},
  cell{4}{4} = {cDeviceToDeviceHeadEYKT},
  hlines,
  vlines,
}
\diagbox{Data Type}{Model} & Simple & LSTM & EYKT & Chance \\
Facial                   &  \DeviceToDeviceFaceSimple  & \DeviceToDeviceFaceLSTM & \DeviceToDeviceFaceEYKT & \DeviceToDeviceFaceChance \\
Eye                    &  \DeviceToDeviceEyeSimple  & \DeviceToDeviceEyeLSTM & \DeviceToDeviceEyeEYKT & \DeviceToDeviceEyeChance \\
Head                   &  \DeviceToDeviceHeadSimple  & \DeviceToDeviceHeadLSTM & \DeviceToDeviceHeadEYKT & \DeviceToDeviceHeadChance        
\end{tblr}
\end{table}

Next, we test if we can recognize participants across different \ac{MR} headsets in Experiment \textbf{E3} (see Table~\ref{tab:DeviceToDevice}). The LHPP split leaves for every participant one headset type for which the model has not seen any data, hence, we simulate that the user switches to a new type of \ac{MR} headset. The best accuracy for facial motion data is achieved by EYKT with 63\%, showing that identifying individuals across headsets is possible, however, at a lower rate than in our baseline E1.

\providecommand{\EmotionRecFaceSimple}{$0.86$}
\definecolor{cEmotionRecFaceSimple}{rgb}{0.636902,0.856542,0.21662}
\providecommand{\EmotionRecFaceLSTM}{$0.86$}
\definecolor{cEmotionRecFaceLSTM}{rgb}{0.636902,0.856542,0.21662}
\providecommand{\EmotionRecFaceEYKT}{$0.86$}
\definecolor{cEmotionRecFaceEYKT}{rgb}{0.636902,0.856542,0.21662}
\providecommand{\EmotionRecFaceChance}{$0.33$}
\providecommand{\EmotionRecHeadSimple}{$0.49$}
\definecolor{cEmotionRecHeadSimple}{rgb}{0.131172,0.555899,0.552459}
\providecommand{\EmotionRecHeadLSTM}{$0.32$}
\definecolor{cEmotionRecHeadLSTM}{rgb}{0.197636,0.391528,0.554969}
\providecommand{\EmotionRecHeadEYKT}{$0.58$}
\definecolor{cEmotionRecHeadEYKT}{rgb}{0.12478,0.640461,0.527068}
\providecommand{\EmotionRecHeadChance}{$0.33$}
\providecommand{\EmotionRecEyeSimple}{$0.45$}
\definecolor{cEmotionRecEyeSimple}{rgb}{0.144759,0.519093,0.556572}
\providecommand{\EmotionRecEyeLSTM}{$0.33$}
\definecolor{cEmotionRecEyeLSTM}{rgb}{0.192357,0.403199,0.555836}
\providecommand{\EmotionRecEyeEYKT}{$0.59$}
\definecolor{cEmotionRecEyeEYKT}{rgb}{0.130067,0.651384,0.521608}
\providecommand{\EmotionRecEyeChance}{$0.33$}

\begin{table}
\caption{Emotion recognition accuracy using a participant-wise split for all headsets}\label{tab:EmotionRec}
\centering
\begin{tblr}{
  cell{2}{2} = {cEmotionRecFaceSimple},
  cell{2}{3} = {cEmotionRecFaceLSTM},
  cell{2}{4} = {cEmotionRecFaceEYKT},
  cell{3}{2} = {cEmotionRecEyeSimple},
  cell{3}{3} = {cEmotionRecEyeLSTM},
  cell{3}{4} = {cEmotionRecEyeEYKT},
  cell{4}{2} = {cEmotionRecHeadSimple},
  cell{4}{3} = {cEmotionRecHeadLSTM},
  cell{4}{4} = {cEmotionRecHeadEYKT},
  hlines,
  vlines,
}
\diagbox{Data Type}{Model} & Simple & LSTM & EYKT & Chance \\
Facial                   &  \EmotionRecFaceSimple  & \EmotionRecFaceLSTM & \EmotionRecFaceEYKT & \EmotionRecFaceChance \\
Eye                    &  \EmotionRecEyeSimple  & \EmotionRecEyeLSTM & \EmotionRecEyeEYKT & \EmotionRecEyeChance \\
Head                   &  \EmotionRecHeadSimple  & \EmotionRecHeadLSTM & \EmotionRecHeadEYKT & \EmotionRecHeadChance        
\end{tblr}
\end{table}

In our Experiment \textbf{E4}, we tested emotion recognition using only emotion tasks (see Table~\ref{tab:EmotionRec}). As expected, facial motion data was the most effective for emotion recognition, with 86\% accuracy. However, eye and head motions also enabled some emotion recognition, with accuracy rates of 59\% and 58\%, respectively. %

\providecommand{\DeviceIDFaceSimple}{$1.0$}
\definecolor{cDeviceIDFaceSimple}{rgb}{0.993248,0.906157,0.143936}
\providecommand{\DeviceIDFaceLSTM}{$1.0$}
\definecolor{cDeviceIDFaceLSTM}{rgb}{0.993248,0.906157,0.143936}
\providecommand{\DeviceIDFaceEYKT}{$1.0$}
\definecolor{cDeviceIDFaceEYKT}{rgb}{0.993248,0.906157,0.143936}
\providecommand{\DeviceIDFaceChance}{$0.48$}
\providecommand{\DeviceIDHeadSimple}{$1.0$}
\definecolor{cDeviceIDHeadSimple}{rgb}{0.993248,0.906157,0.143936}
\providecommand{\DeviceIDHeadLSTM}{$1.0$}
\definecolor{cDeviceIDHeadLSTM}{rgb}{0.993248,0.906157,0.143936}
\providecommand{\DeviceIDHeadEYKT}{$1.0$}
\definecolor{cDeviceIDHeadEYKT}{rgb}{0.993248,0.906157,0.143936}
\providecommand{\DeviceIDHeadChance}{$0.48$}
\providecommand{\DeviceIDEyeSimple}{$1.0$}
\definecolor{cDeviceIDEyeSimple}{rgb}{0.993248,0.906157,0.143936}
\providecommand{\DeviceIDEyeLSTM}{$1.0$}
\definecolor{cDeviceIDEyeLSTM}{rgb}{0.993248,0.906157,0.143936}
\providecommand{\DeviceIDEyeEYKT}{$1.0$}
\definecolor{cDeviceIDEyeEYKT}{rgb}{0.993248,0.906157,0.143936}
\providecommand{\DeviceIDEyeChance}{$0.48$}

\begin{table}
\caption{Device type recognition accuracy using a participant-wise split for all headsets}\label{tab:DeviceID}
\centering
\begin{tblr}{
  cell{2}{2} = {cDeviceIDFaceSimple},
  cell{2}{3} = {cDeviceIDFaceLSTM},
  cell{2}{4} = {cDeviceIDFaceEYKT},
  cell{3}{2} = {cDeviceIDEyeSimple},
  cell{3}{3} = {cDeviceIDEyeLSTM},
  cell{3}{4} = {cDeviceIDEyeEYKT},
  cell{4}{2} = {cDeviceIDHeadSimple},
  cell{4}{3} = {cDeviceIDHeadLSTM},
  cell{4}{4} = {cDeviceIDHeadEYKT},
  hlines,
  vlines,
}
\diagbox{Data Type}{Model} & Simple & LSTM & EYKT & Chance \\
Facial                   &  \DeviceIDFaceSimple  & \DeviceIDFaceLSTM & \DeviceIDFaceEYKT & \DeviceIDFaceChance \\
Eye                    &  \DeviceIDEyeSimple  & \DeviceIDEyeLSTM & \DeviceIDEyeEYKT & \DeviceIDEyeChance \\
Head                   &  \DeviceIDHeadSimple  & \DeviceIDHeadLSTM & \DeviceIDHeadEYKT & \DeviceIDHeadChance        
\end{tblr}
\end{table}

Since we have multiple devices, we tested whether we could identify which headset was used to record the data for Experiment \textbf{E5} (see Table~\ref{tab:DeviceID}). Unsurprisingly, we can achieve 100\% recognition accuracy for all data types. %

\providecommand{\TextRecFaceSimple}{$0.78$}
\definecolor{cTextRecFaceSimple}{rgb}{0.430983,0.808473,0.346476}
\providecommand{\TextRecFaceLSTM}{$0.89$}
\definecolor{cTextRecFaceLSTM}{rgb}{0.709898,0.868751,0.169257}
\providecommand{\TextRecFaceEYKT}{$0.96$}
\definecolor{cTextRecFaceEYKT}{rgb}{0.89632,0.893616,0.096335}
\providecommand{\TextRecFaceChance}{$0.17$}
\providecommand{\TextRecHeadSimple}{\textcolor{white}{$0.17$}}
\definecolor{cTextRecHeadSimple}{rgb}{0.26658,0.228262,0.514349}
\providecommand{\TextRecHeadLSTM}{\textcolor{white}{$0.17$}}
\definecolor{cTextRecHeadLSTM}{rgb}{0.26658,0.228262,0.514349}
\providecommand{\TextRecHeadEYKT}{$0.47$}
\definecolor{cTextRecHeadEYKT}{rgb}{0.13777,0.537492,0.554906}
\providecommand{\TextRecHeadChance}{$0.17$}
\providecommand{\TextRecEyeSimple}{$0.56$}
\definecolor{cTextRecEyeSimple}{rgb}{0.120081,0.622161,0.534946}
\providecommand{\TextRecEyeLSTM}{$0.6$}
\definecolor{cTextRecEyeLSTM}{rgb}{0.134692,0.658636,0.517649}
\providecommand{\TextRecEyeEYKT}{$0.68$}
\definecolor{cTextRecEyeEYKT}{rgb}{0.232815,0.732247,0.459277}
\providecommand{\TextRecEyeChance}{$0.17$}

\begin{table}
\caption{Verbal task recognition accuracy using a participant-wise split for all headsets}\label{tab:TextRec}
\centering
\begin{tblr}{
  cell{2}{2} = {cTextRecFaceSimple},
  cell{2}{3} = {cTextRecFaceLSTM},
  cell{2}{4} = {cTextRecFaceEYKT},
  cell{3}{2} = {cTextRecEyeSimple},
  cell{3}{3} = {cTextRecEyeLSTM},
  cell{3}{4} = {cTextRecEyeEYKT},
  cell{4}{2} = {cTextRecHeadSimple},
  cell{4}{3} = {cTextRecHeadLSTM},
  cell{4}{4} = {cTextRecHeadEYKT},
  hlines,
  vlines,
}
\diagbox{Data Type}{Model} & Simple & LSTM & EYKT & Chance \\
Facial                   &  \TextRecFaceSimple  & \TextRecFaceLSTM & \TextRecFaceEYKT & \TextRecFaceChance \\
Eye                    &  \TextRecEyeSimple  & \TextRecEyeLSTM & \TextRecEyeEYKT & \TextRecEyeChance \\
Head                   &  \TextRecHeadSimple  & \TextRecHeadLSTM & \TextRecHeadEYKT & \TextRecHeadChance        
\end{tblr}
\end{table}

In addition to recognizing emotions, we test whether the text task can be identified from the recorded data (see Table~\ref{tab:TextRec}). The best recognition accuracy of 96\% is again achieved using facial motion data.

\providecommand{\EnglishFaceSimple}{$0.72$}
\definecolor{cEnglishFaceSimple}{rgb}{0.304148,0.764704,0.419943}
\providecommand{\EnglishFaceLSTM}{$0.73$}
\definecolor{cEnglishFaceLSTM}{rgb}{0.319809,0.770914,0.411152}
\providecommand{\EnglishFaceEYKT}{$0.69$}
\definecolor{cEnglishFaceEYKT}{rgb}{0.24607,0.73891,0.452024}
\providecommand{\EnglishFaceChance}{$0.73$}
\providecommand{\EnglishHeadSimple}{$0.73$}
\definecolor{cEnglishHeadSimple}{rgb}{0.319809,0.770914,0.411152}
\providecommand{\EnglishHeadLSTM}{$0.73$}
\definecolor{cEnglishHeadLSTM}{rgb}{0.319809,0.770914,0.411152}
\providecommand{\EnglishHeadEYKT}{$0.68$}
\definecolor{cEnglishHeadEYKT}{rgb}{0.232815,0.732247,0.459277}
\providecommand{\EnglishHeadChance}{$0.73$}
\providecommand{\EnglishEyeSimple}{$0.73$}
\definecolor{cEnglishEyeSimple}{rgb}{0.319809,0.770914,0.411152}
\providecommand{\EnglishEyeLSTM}{$0.73$}
\definecolor{cEnglishEyeLSTM}{rgb}{0.319809,0.770914,0.411152}
\providecommand{\EnglishEyeEYKT}{$0.71$}
\definecolor{cEnglishEyeEYKT}{rgb}{0.281477,0.755203,0.432552}
\providecommand{\EnglishEyeChance}{$0.73$}

\begin{table}
\caption{English level recognition accuracy using a participant-wise split for all headsets}\label{tab:English}
\centering
\begin{tblr}{
  cell{2}{2} = {cEnglishFaceSimple},
  cell{2}{3} = {cEnglishFaceLSTM},
  cell{2}{4} = {cEnglishFaceEYKT},
  cell{3}{2} = {cEnglishEyeSimple},
  cell{3}{3} = {cEnglishEyeLSTM},
  cell{3}{4} = {cEnglishEyeEYKT},
  cell{4}{2} = {cEnglishHeadSimple},
  cell{4}{3} = {cEnglishHeadLSTM},
  cell{4}{4} = {cEnglishHeadEYKT},
  hlines,
  vlines,
}
\diagbox{Data Type}{Model} & Simple & LSTM & EYKT & Chance \\
Facial                   &  \EnglishFaceSimple  & \EnglishFaceLSTM & \EnglishFaceEYKT & \EnglishFaceChance \\
Eye                    &  \EnglishEyeSimple  & \EnglishEyeLSTM & \EnglishEyeEYKT & \EnglishEyeChance \\
Head                   &  \EnglishHeadSimple  & \EnglishHeadLSTM & \EnglishHeadEYKT & \EnglishHeadChance        
\end{tblr}
\end{table}

\providecommand{\PersonalityFaceSimple}{$0.43$}
\definecolor{cPersonalityFaceSimple}{rgb}{0.151918,0.500685,0.557587}
\providecommand{\PersonalityFaceLSTM}{$0.49$}
\definecolor{cPersonalityFaceLSTM}{rgb}{0.131172,0.555899,0.552459}
\providecommand{\PersonalityFaceEYKT}{$0.41$}
\definecolor{cPersonalityFaceEYKT}{rgb}{0.160665,0.47854,0.558115}
\providecommand{\PersonalityFaceChance}{$0.49$}
\providecommand{\PersonalityHeadSimple}{$0.49$}
\definecolor{cPersonalityHeadSimple}{rgb}{0.131172,0.555899,0.552459}
\providecommand{\PersonalityHeadLSTM}{$0.49$}
\definecolor{cPersonalityHeadLSTM}{rgb}{0.131172,0.555899,0.552459}
\providecommand{\PersonalityHeadEYKT}{$0.38$}
\definecolor{cPersonalityHeadEYKT}{rgb}{0.171176,0.45253,0.557965}
\providecommand{\PersonalityHeadChance}{$0.49$}
\providecommand{\PersonalityEyeSimple}{$0.44$}
\definecolor{cPersonalityEyeSimple}{rgb}{0.149039,0.508051,0.55725}
\providecommand{\PersonalityEyeLSTM}{$0.47$}
\definecolor{cPersonalityEyeLSTM}{rgb}{0.13777,0.537492,0.554906}
\providecommand{\PersonalityEyeEYKT}{$0.41$}
\definecolor{cPersonalityEyeEYKT}{rgb}{0.160665,0.47854,0.558115}
\providecommand{\PersonalityEyeChance}{$0.49$}

\begin{table}
\caption{Personality recognition accuracy using a participant-wise split for all headsets}\label{tab:Personality}
\centering
\begin{tblr}{
  cell{2}{2} = {cPersonalityFaceSimple},
  cell{2}{3} = {cPersonalityFaceLSTM},
  cell{2}{4} = {cPersonalityFaceEYKT},
  cell{3}{2} = {cPersonalityEyeSimple},
  cell{3}{3} = {cPersonalityEyeLSTM},
  cell{3}{4} = {cPersonalityEyeEYKT},
  cell{4}{2} = {cPersonalityHeadSimple},
  cell{4}{3} = {cPersonalityHeadLSTM},
  cell{4}{4} = {cPersonalityHeadEYKT},
  hlines,
  vlines,
}
\diagbox{Data Type}{Model} & Simple & LSTM & EYKT & Chance \\
Facial                   &  \PersonalityFaceSimple  & \PersonalityFaceLSTM & \PersonalityFaceEYKT & \PersonalityFaceChance \\
Eye                    &  \PersonalityEyeSimple  & \PersonalityEyeLSTM & \PersonalityEyeEYKT & \PersonalityEyeChance \\
Head                   &  \PersonalityHeadSimple  & \PersonalityHeadLSTM & \PersonalityHeadEYKT & \PersonalityHeadChance        
\end{tblr}
\end{table}

\providecommand{\SexRecFaceSimple}{$0.65$}
\definecolor{cSexRecFaceSimple}{rgb}{0.185783,0.704891,0.485273}
\providecommand{\SexRecFaceLSTM}{$0.81$}
\definecolor{cSexRecFaceLSTM}{rgb}{0.506271,0.828786,0.300362}
\providecommand{\SexRecFaceEYKT}{$0.67$}
\definecolor{cSexRecFaceEYKT}{rgb}{0.214,0.722114,0.469588}
\providecommand{\SexRecFaceChance}{$0.81$}
\providecommand{\SexRecHeadSimple}{$0.73$}
\definecolor{cSexRecHeadSimple}{rgb}{0.319809,0.770914,0.411152}
\providecommand{\SexRecHeadLSTM}{$0.81$}
\definecolor{cSexRecHeadLSTM}{rgb}{0.506271,0.828786,0.300362}
\providecommand{\SexRecHeadEYKT}{$0.56$}
\definecolor{cSexRecHeadEYKT}{rgb}{0.120081,0.622161,0.534946}
\providecommand{\SexRecHeadChance}{$0.81$}
\providecommand{\SexRecEyeSimple}{$0.72$}
\definecolor{cSexRecEyeSimple}{rgb}{0.304148,0.764704,0.419943}
\providecommand{\SexRecEyeLSTM}{$0.81$}
\definecolor{cSexRecEyeLSTM}{rgb}{0.506271,0.828786,0.300362}
\providecommand{\SexRecEyeEYKT}{$0.58$}
\definecolor{cSexRecEyeEYKT}{rgb}{0.12478,0.640461,0.527068}
\providecommand{\SexRecEyeChance}{$0.81$}

\begin{table}
\caption{Sex recognition accuracy using a participant-wise split for all headsets}\label{tab:SexRec}
\centering
\begin{tblr}{
  cell{2}{2} = {cSexRecFaceSimple},
  cell{2}{3} = {cSexRecFaceLSTM},
  cell{2}{4} = {cSexRecFaceEYKT},
  cell{3}{2} = {cSexRecEyeSimple},
  cell{3}{3} = {cSexRecEyeLSTM},
  cell{3}{4} = {cSexRecEyeEYKT},
  cell{4}{2} = {cSexRecHeadSimple},
  cell{4}{3} = {cSexRecHeadLSTM},
  cell{4}{4} = {cSexRecHeadEYKT},
  hlines,
  vlines,
}
\diagbox{Data Type}{Model} & Simple & LSTM & EYKT & Chance \\
Facial                   &  \SexRecFaceSimple  & \SexRecFaceLSTM & \SexRecFaceEYKT & \SexRecFaceChance \\
Eye                    &  \SexRecEyeSimple  & \SexRecEyeLSTM & \SexRecEyeEYKT & \SexRecEyeChance \\
Head                   &  \SexRecHeadSimple  & \SexRecHeadLSTM & \SexRecHeadEYKT & \SexRecHeadChance        
\end{tblr}
\end{table}

We examine the results of the attribute inferences tested in Experiment \textbf{E6}. Table~\ref{tab:English} shows the results for English level recognition, and Table~\ref{tab:Personality} shows the results for classifying whether someone is an ambivert, extrovert, or introvert. For both attributes, the results are close to the level of chance, so we do not believe they can be inferred from the data. For sex recognition, shown in Table~\ref{tab:SexRec}, there appears to be some information which can be extracted. Since the EYKT model achieved significantly less than chance level, and with only two classes (everyone identified as either male or female) in the dataset, we can simply invert the labeling.

\providecommand{\OnlyTextFaceSimple}{$0.72$}
\definecolor{cOnlyTextFaceSimple}{rgb}{0.304148,0.764704,0.419943}
\providecommand{\OnlyTextFaceLSTM}{$0.65$}
\definecolor{cOnlyTextFaceLSTM}{rgb}{0.185783,0.704891,0.485273}
\providecommand{\OnlyTextFaceEYKT}{$0.93$}
\definecolor{cOnlyTextFaceEYKT}{rgb}{0.82494,0.88472,0.106217}
\providecommand{\OnlyTextFaceChance}{$0.01$}
\providecommand{\OnlyTextHeadSimple}{$0.79$}
\definecolor{cOnlyTextHeadSimple}{rgb}{0.458674,0.816363,0.329727}
\providecommand{\OnlyTextHeadLSTM}{$0.77$}
\definecolor{cOnlyTextHeadLSTM}{rgb}{0.412913,0.803041,0.357269}
\providecommand{\OnlyTextHeadEYKT}{$0.95$}
\definecolor{cOnlyTextHeadEYKT}{rgb}{0.876168,0.891125,0.09525}
\providecommand{\OnlyTextHeadChance}{$0.01$}
\providecommand{\OnlyTextEyeSimple}{$0.52$}
\definecolor{cOnlyTextEyeSimple}{rgb}{0.122606,0.585371,0.546557}
\providecommand{\OnlyTextEyeLSTM}{$0.56$}
\definecolor{cOnlyTextEyeLSTM}{rgb}{0.120081,0.622161,0.534946}
\providecommand{\OnlyTextEyeEYKT}{$0.83$}
\definecolor{cOnlyTextEyeEYKT}{rgb}{0.555484,0.840254,0.269281}
\providecommand{\OnlyTextEyeChance}{$0.01$}

\providecommand{\OnlyAnimationFaceSimple}{$0.63$}
\definecolor{cOnlyAnimationFaceSimple}{rgb}{0.162016,0.687316,0.499129}
\providecommand{\OnlyAnimationFaceLSTM}{$0.47$}
\definecolor{cOnlyAnimationFaceLSTM}{rgb}{0.13777,0.537492,0.554906}
\providecommand{\OnlyAnimationFaceEYKT}{$0.8$}
\definecolor{cOnlyAnimationFaceEYKT}{rgb}{0.477504,0.821444,0.318195}
\providecommand{\OnlyAnimationFaceChance}{$0.01$}
\providecommand{\OnlyAnimationHeadSimple}{$0.64$}
\definecolor{cOnlyAnimationHeadSimple}{rgb}{0.170948,0.694384,0.493803}
\providecommand{\OnlyAnimationHeadLSTM}{$0.41$}
\definecolor{cOnlyAnimationHeadLSTM}{rgb}{0.160665,0.47854,0.558115}
\providecommand{\OnlyAnimationHeadEYKT}{$0.89$}
\definecolor{cOnlyAnimationHeadEYKT}{rgb}{0.709898,0.868751,0.169257}
\providecommand{\OnlyAnimationHeadChance}{$0.01$}
\providecommand{\OnlyAnimationEyeSimple}{$0.38$}
\definecolor{cOnlyAnimationEyeSimple}{rgb}{0.171176,0.45253,0.557965}
\providecommand{\OnlyAnimationEyeLSTM}{$0.33$}
\definecolor{cOnlyAnimationEyeLSTM}{rgb}{0.192357,0.403199,0.555836}
\providecommand{\OnlyAnimationEyeEYKT}{$0.75$}
\definecolor{cOnlyAnimationEyeEYKT}{rgb}{0.369214,0.788888,0.382914}
\providecommand{\OnlyAnimationEyeChance}{$0.01$}

\begin{table}
\caption{Identification accuracy using a random split for only verbal tasks for all headsets}\label{tab:OnlyText}
\centering
\begin{tblr}{
  cell{2}{2} = {cOnlyTextFaceSimple},
  cell{2}{3} = {cOnlyTextFaceLSTM},
  cell{2}{4} = {cOnlyTextFaceEYKT},
  cell{3}{2} = {cOnlyTextEyeSimple},
  cell{3}{3} = {cOnlyTextEyeLSTM},
  cell{3}{4} = {cOnlyTextEyeEYKT},
  cell{4}{2} = {cOnlyTextHeadSimple},
  cell{4}{3} = {cOnlyTextHeadLSTM},
  cell{4}{4} = {cOnlyTextHeadEYKT},
  hlines,
  vlines,
}
\diagbox{Data Type}{Model} & Simple & LSTM & EYKT & Chance \\
Face                   &  \OnlyTextFaceSimple  & \OnlyTextFaceLSTM & \OnlyTextFaceEYKT & \OnlyTextFaceChance \\
Eye                    &  \OnlyTextEyeSimple  & \OnlyTextEyeLSTM & \OnlyTextEyeEYKT & \OnlyTextEyeChance \\
Head                   &  \OnlyTextHeadSimple  & \OnlyTextHeadLSTM & \OnlyTextHeadEYKT & \OnlyTextHeadChance        
\end{tblr}
\end{table}

\begin{table}
\caption{Identification accuracy using a random split for only verbal tasks for all headsets}\label{tab:OnlyAnimation}
\centering
\begin{tblr}{
  cell{2}{2} = {cOnlyAnimationFaceSimple},
  cell{2}{3} = {cOnlyAnimationFaceLSTM},
  cell{2}{4} = {cOnlyAnimationFaceEYKT},
  cell{3}{2} = {cOnlyAnimationEyeSimple},
  cell{3}{3} = {cOnlyAnimationEyeLSTM},
  cell{3}{4} = {cOnlyAnimationEyeEYKT},
  cell{4}{2} = {cOnlyAnimationHeadSimple},
  cell{4}{3} = {cOnlyAnimationHeadLSTM},
  cell{4}{4} = {cOnlyAnimationHeadEYKT},
  hlines,
  vlines,
}
\diagbox{Data Type}{Model} & Simple & LSTM & EYKT & Chance \\
Facial                   &  \OnlyAnimationFaceSimple  & \OnlyAnimationFaceLSTM & \OnlyAnimationFaceEYKT & \OnlyAnimationFaceChance \\
Eye                    &  \OnlyAnimationEyeSimple  & \OnlyAnimationEyeLSTM & \OnlyAnimationEyeEYKT & \OnlyAnimationEyeChance \\
Head                   &  \OnlyAnimationHeadSimple  & \OnlyAnimationHeadLSTM & \OnlyAnimationHeadEYKT & \OnlyAnimationHeadChance        
\end{tblr}
\end{table}

To better understand which task type is better for identifying individuals, we ran identification Experiment \textbf{E7} on only the verbal and non-verbal tasks. See Tables \ref{tab:OnlyText} and \ref{tab:OnlyAnimation} for a comparison. Our results show that verbal tasks perform better than non-verbal tasks. However, it also does not appear that a reliable sex recognition can be implemented with facial motion data for now.

\providecommand{\DataTypeCombFaceEyeHMDSimple}{$0.89$}
\definecolor{cDataTypeCombFaceEyeHMDSimple}{rgb}{0.709898,0.868751,0.169257}
\providecommand{\DataTypeCombFaceEyeHMDLSTM}{$0.83$}
\definecolor{cDataTypeCombFaceEyeHMDLSTM}{rgb}{0.555484,0.840254,0.269281}
\providecommand{\DataTypeCombFaceEyeHMDEYKT}{$0.99$}
\definecolor{cDataTypeCombFaceEyeHMDEYKT}{rgb}{0.974417,0.90359,0.130215}
\providecommand{\DataTypeCombFaceEyeHMDChance}{$0.01$}

\begin{table}
\caption{Identification accuracy using a random split for all headsets}\label{tab:DataTypeComb}
\centering
\begin{tblr}{
  cell{2}{2} = {cDataTypeCombFaceEyeHMDSimple},
  cell{2}{3} = {cDataTypeCombFaceEyeHMDLSTM},
  cell{2}{4} = {cDataTypeCombFaceEyeHMDEYKT},
  hlines,
  vlines,
}
\diagbox{Data Type}{Model} & Simple & LSTM & EYKT & Chance \\
Facial + Eye + Head &  \DataTypeCombFaceEyeHMDSimple  & \DataTypeCombFaceEyeHMDLSTM & \DataTypeCombFaceEyeHMDEYKT & \DataTypeCombFaceEyeHMDChance
\end{tblr}
\end{table}

Lastly, we further investigated the identification potential of the data we collected. In Experiment \textbf{E8} (see Table~\ref{tab:DataTypeComb}), we tested the identification accuracy using all data types simultaneously. We found that combining the three data types increased identification accuracy to 99\%, thereby outperforming the best all-headset result from Experiment E1 (see Table~\ref{tab:Baseline}).

\subsection{Summary of Results}

\begin{itemize}
    \item We are able to show that persons can be identified from their facial motions.
    \item The identification across different sessions is possible, however, the achieved accuracy is not on a level that is usable for any real-world system at the moment.
    \item We are able to show identification across different \ac{MR} headset types.
    \item We are able to infer the displayed emotion, spoken text, and used \ac{MR} headset.
    \item We are not able to infer the personality and spoken language of a person.
    \item For sex recognition, we see some indication that some sex-related information is contained in facial motion data.
\end{itemize}

\section{Discussion}\label{sec:discussion}

Facial motion data is a behavioral biometric factor that can be used for identification, so it should be treated as such when sharing it online. However, our results indicate that facial motion might not be stable enough to reliable identify individuals over long periods of time. Only larger studies with longer intervals between sessions can determine whether facial motion data poses a long-term privacy threat to individuals. We expect \ac{MR} headsets to improve their ability to record facial motion data in the future, so we also expect privacy problems with facial motion data to increase.

Our text recognition results show that we can infer which text was spoken. This suggests that lip reading may be possible using facial motion data. Uncareful sharing of facial motion data might lead to the revelation of the content of private conversations, for example, when a person has muted themselves but is still sharing their facial motion data.

When we compare our eye gaze and head motion results to those of previous studies, such as GazebaseVR~\cite{Lohr_2023} for eye gaze data and Nair et al.~\cite{nair2023unique} for VR data, we find that our identification results are are not as good, especially when considering multiple sessions. We believe this is because the tasks in our dataset are designed primarily to capture facial motion data. For example, GazebaseVR uses specific eye-tracking tasks, such as following a dot with one's eyes or reading tasks. In contrast, we only record data after participants read the tasks and push the button to start recording; therefore, we do not expect much eye motion during recording. Additionally, none of our tasks require head motion, so little variance is expected.

\section{Conclusion}\label{sec:conclusion}

In this paper, we present the first large-scale dataset of abstract face motions captured using \ac{MR} headsets. The dataset contains multiple sessions, and each participant is recorded using multiple headset types. Using this dataset, we demonstrate that facial motion data is a privacy-sensitive behavioral biometric factor that can be used to identify individuals with up to 98\% not considering sessions and 43\% when considering sessions. Furthermore, we demonstrate that individuals can be identified even when using a new type of \ac{MR} headset that the attacker has not seen before.

\ifCLASSOPTIONcompsoc
  \section*{Acknowledgments}
\ifanon
\else
Funded by the German Research Foundation (DFG, Deutsche Forschungsgemeinschaft) as part of Germany’s Excellence Strategy – EXC 2050/1 – Project ID 390696704 – Cluster of Excellence “Centre for Tactile Internet with Human-in-the-Loop” (CeTI) of Technische Universität Dresden; Funded by Helmholtz Association, topic ``46.23 Engineering Secure Systems''. The studies were conducted in the Karlsruhe Decision\&Design Lab (KD²Lab), an experimental laboratory funded by the DFG and the Karlsruhe Institute of Technology (INST\_12138411-1\_FUGG).
\fi
\else
  \section*{Acknowledgment}
\fi

\bibliographystyle{ieeetr}
\bibliography{facial_motion}

\end{document}